\documentclass{article}\usepackage[letterpaper]{geometry}
\usepackage{natbib}
\usepackage{amsmath,amssymb,amsthm}
\usepackage{tabularx,colortbl}
\usepackage{multirow}
\usepackage{times}
\usepackage{graphicx}
\usepackage{epsfig}
\usepackage{caption}

\title{Bayesian Modeling of Inconsistent Plastic Response due to Material Variability}
\author{
  F. Rizzi\thanks{Sandia National Laboratories, P.O. Box 969, Livermore, CA 94551, USA.},\
  M. Khalil,\
  R.E. Jones\thanks{Corresponding Author Email: rjones@sandia.gov.},\
  J.A. Templeton,\
  J.T. Ostien,\
  B.L. Boyce
}
\date{}

\newcommand{\bit}{\begin{itemize}}
\newcommand{\eit}{\end{itemize}}
\newcommand{\benu}{\begin{enumerate}}
\newcommand{\eenu}{\end{enumerate}}
\newcommand{\be}{\begin{equation}}
\newcommand{\ee}{\end{equation}}
\newcommand{\bea}{\begin{eqnarray}}
\newcommand{\eea}{\end{eqnarray}}
\newcommand{\bean}{\begin{eqnarray*}}
\newcommand{\eean}{\end{eqnarray*}}
\newcommand{\ben}{\begin{equation*}}		
\newcommand{\een}{\end{equation*}}

\newlength{\figwidth}
\newlength{\figwidthtwo}
\newlength{\figwidththree}
\newcommand{\figref}[1]{Fig. \ref{#1}}

\newcommand{\fref}[1]{Fig.\,\ref{#1}}
\newcommand{\tref}[1]{Table\,\ref{#1}}
\newcommand{\eref}[1]{Eq.\,(\ref{#1})}

\newcommand{\sref}[1]{Sec.\!~\ref{#1}}

\newcommand{\cref}[1]{Ref.\,\cite{#1}}

\newcommand{\cf}{{\it cf.}\! }

\newcommand{\ie}{{\it i.e.}\! }

\newcommand{\etal}{{\it et al.}\! }

\newcommand{\evidence}{\eta}

\newcommand{\prob}{p} 
\newcommand{\strain}{\varepsilon}
\newcommand{\young}{E}
\newcommand{\yield}{Y}
\newcommand{\hard}{H}
\newcommand{\satmod}{K}
\newcommand{\satexp}{B}
\newcommand{\stress}{{\sigma}}

\newcommand{\Dc}{\mathcal{D}}

\newcommand{\bb}{\mathbf{b}}

\renewcommand{\sb}{\mathbf{s}}
\newcommand{\alphab}{{\boldsymbol{\alpha}}}

\newcommand{\xib}{{\boldsymbol{\xi}}}
\newcommand{\thetab}{{\boldsymbol{\theta}}}

\newcommand{\taub}{{\boldsymbol{\tau}}}

\newcommand{\Fb}{\mathbf{F}}

\newcommand{\Ib}{\mathbf{I}}

\newcommand{\tr}{\operatorname{tr}}

\newcommand{\dev}{\operatorname{dev}}

\usepackage{color}
\usepackage[normalem]{ulem}

\begin{document}
\maketitle

\abstract{
The advent of fabrication techniques such as additive manufacturing has focused attention on the considerable variability of material response due to defects and other microstructural aspects.
This variability motivates the development of an enhanced design methodology that incorporates inherent material variability to provide robust predictions of performance.
In this work, we develop plasticity models capable of representing the distribution of mechanical responses observed in experiments using traditional plasticity models of the mean response and recently developed uncertainty quantification (UQ) techniques.
We demonstrate that the new method provides predictive realizations that are superior to more traditional ones, and how these UQ techniques can be used in model selection and assessing the quality of calibrated physical parameters.
}

\section{Introduction} \label{sec:introduction}

Variability of material response due to defects and other microstructural aspects has been well-known for some time [\cite{hill1963elastic, nemat1999averaging, mcdowell2011representation, mandadapu2012homogenization}].
In many engineering applications inherent material variability has been insignificant, and traditionally the design process is based on the mean or lower-bound response of the chosen materials.
Material failure is a notable exception since it is particularly sensitive to outliers in the distributions of microstructural features [\cite{dingreville2010effect, battaile2015crystal, emery2015predicting}].
Even in cases of material failure, traditional engineering design approaches have been able to successfully ignore material variability through the use of empirical safety factors.
As the engineering community moves towards more physically-realistic and efficient designs while maintaining needed safety, it is necessary to replace these empirical safety factors with confidence bounds that  account for actual material variability.

Currently, additive manufacturing (AM) is of particular technological interest and provides strong motivation to not only model the mean response of materials but also their intrinsic variability.
Additive manufacturing has the distinct advantages of being able to fabricate complex geometries and accelerate the design-build-test cycle through rapid prototyping [\cite{frazier2014metal}]; however, currently, fabrication with this technique suffers from variability in mechanical response due to various sources, including defects imbued by the process, the formation of residual stresses, and geometric variation in the printed parts.
As an example, high throughput tensile data from Boyce \etal [\cite{boyce2017extreme}] clearly shows pronounced variability in the resultant yield and hardening.

The current state of AM technology, and other manufacturing methods with intrinsic variability, e.g. nano- and bio-based, would clearly benefit from an enhanced design methodology accounting for this variability in order to meet performance thresholds with high confidence. 
In this work, we leverage tools from uncertainty quantification (UQ) [\cite{le2010spectral, xiu2010numerical, smith2013uncertainty}] to provide material variability models, realizations, and, ultimately, robust performance predictions.

It is well-known that any model is an approximation of the physical response of a real system. 
Typically, models are characterized by many parameters, and thus appropriately tuning them becomes a key step toward reliable predictions.
The most common approach to model calibration is least-squares regression which yields a deterministic result appropriate for design to the mean.
Bayesian inference methods provide a more general framework for model calibration and parameter estimation by providing a robust framework for handling multiple sources of calibration information as well as a full joint probability density on the target parameters.
Traditionally, Bayesian techniques have been applied in conjunction with additive noise models that are appropriate for modeling external, uncorrelated influences on observed responses.
Recently, a novel technique has been developed to embed the modeled stochasticity in distributions on the physical parameters of the model itself [\cite{Sargsyan:2015}], and in this work we adapt it to model the inherent variability of an AM metal [\cite{boyce2017extreme}].
This is not the only method available in this emerging field of probabilistic modeling of physical processes for engineering applications.
There are commonalities between many of the methods.
Notably, the work of Emery \etal [\cite{emery2015predicting}] applied the stochastic reduced order model (SROM) technique [\cite{field2015efficacy}] to weld failure.
The SROM technique has many of the basic components of the embedded noise model: a surrogate model of the response to physical parameters, a means of propagating distributions of parameters with Monte Carlo (MC) sampling and computing realistic realizations of the predicted response.

In practice, many plausible models exist to represent the trends in the noisy observational data.
Typically, these models aim to capture different aspects of the relevant physical phenomenon or use different paradigms to represent similar aspects.
Some of the available models may be overly-complex representations of the system response in relation to the available observations.
This issue is amplified by the use of models to capture the modeling error in addition to the mean trends.
In this context, the optimal model can be obtained using Bayesian model selection, where optimality is measured in terms of data-fit and model simplicity ~[\cite{Beck2010a,Kass1995,Berger:1996a,Verdinelli:1996}].

In \sref{sec:experiment} of this work, we describe the selected experimental dataset [\cite{boyce2017extreme}] that motivates this effort and provides calibration data. This deep dataset provides real-world relevance that a synthetic dataset would not; however, we apply some pre-processing and simplifying assumptions to facilitate the task of developing the methodology. In \sref{sec:theory}, we review the basic plasticity theory that provides the basis for the material variability models developed in \sref{sec:method}. In \sref{sec:method}, we present the Bayesian framework for model calibration and selection, including the formulation of the likelihood functions that are required in this context. In particular, we adapt both the traditional additive error [\cite{kennedy2001bayesian}] and a more novel embedded error [\cite{Sargsyan:2015}] model. In \sref{sec:results}, we provide the results relating to surrogate model construction, parameter estimation and model selection with the available experimental data. In \sref{sec:conclusion}, we emphasize the innovations of the proposed approach to modeling the mechanical response to microstructural material variability.

\section{Experimental Data} \label{sec:experiment}

\begin{figure}[!tb]
	\captionsetup[subfigure]{justification=centering}
\centering
{\setlength{\tabcolsep}{0mm} \begin{tabular}{c}
{\includegraphics[width=0.4\textwidth]{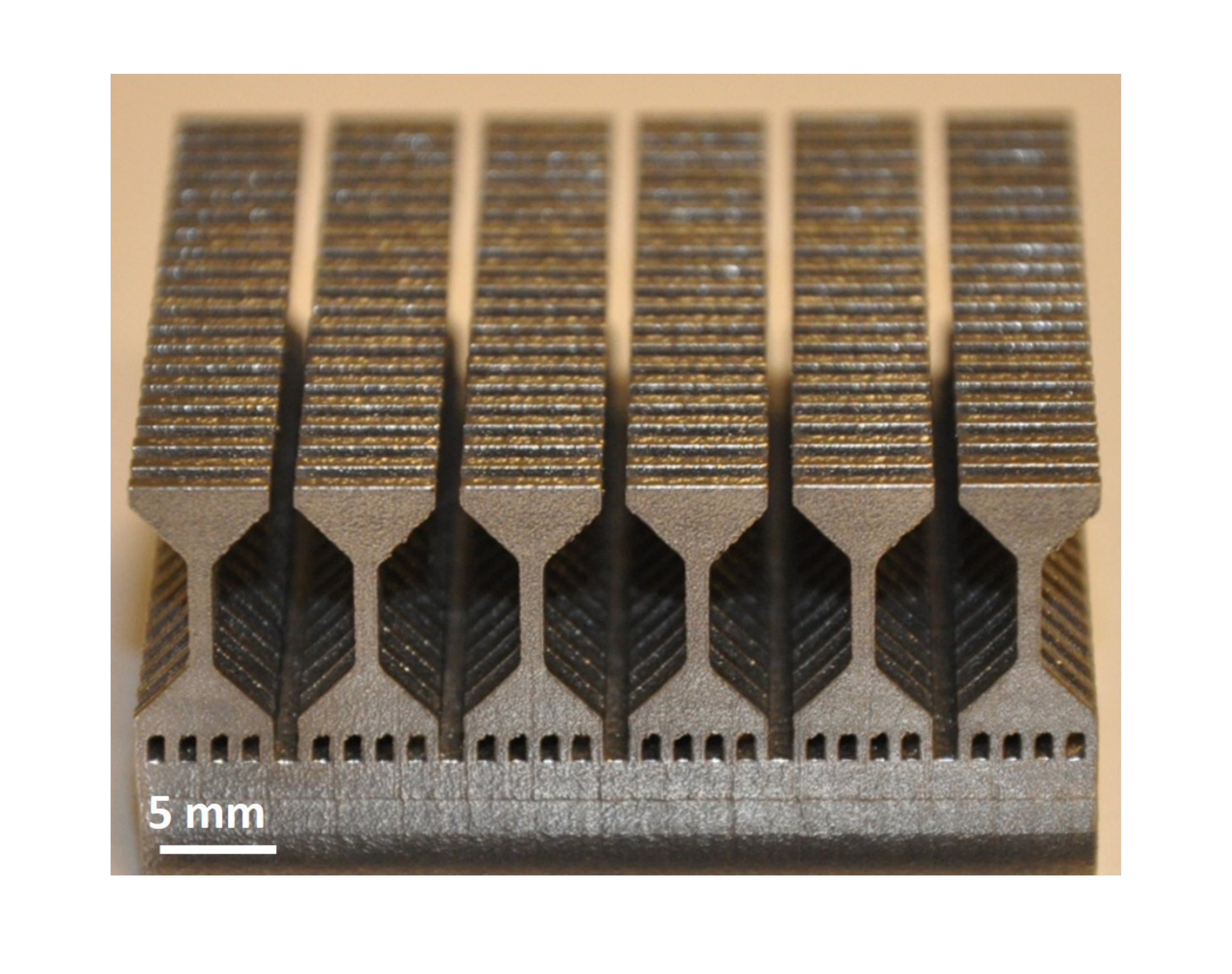}} \\
(a) \\
\end{tabular}
\begin{tabular}{c}
{\includegraphics[width=0.4\textwidth]{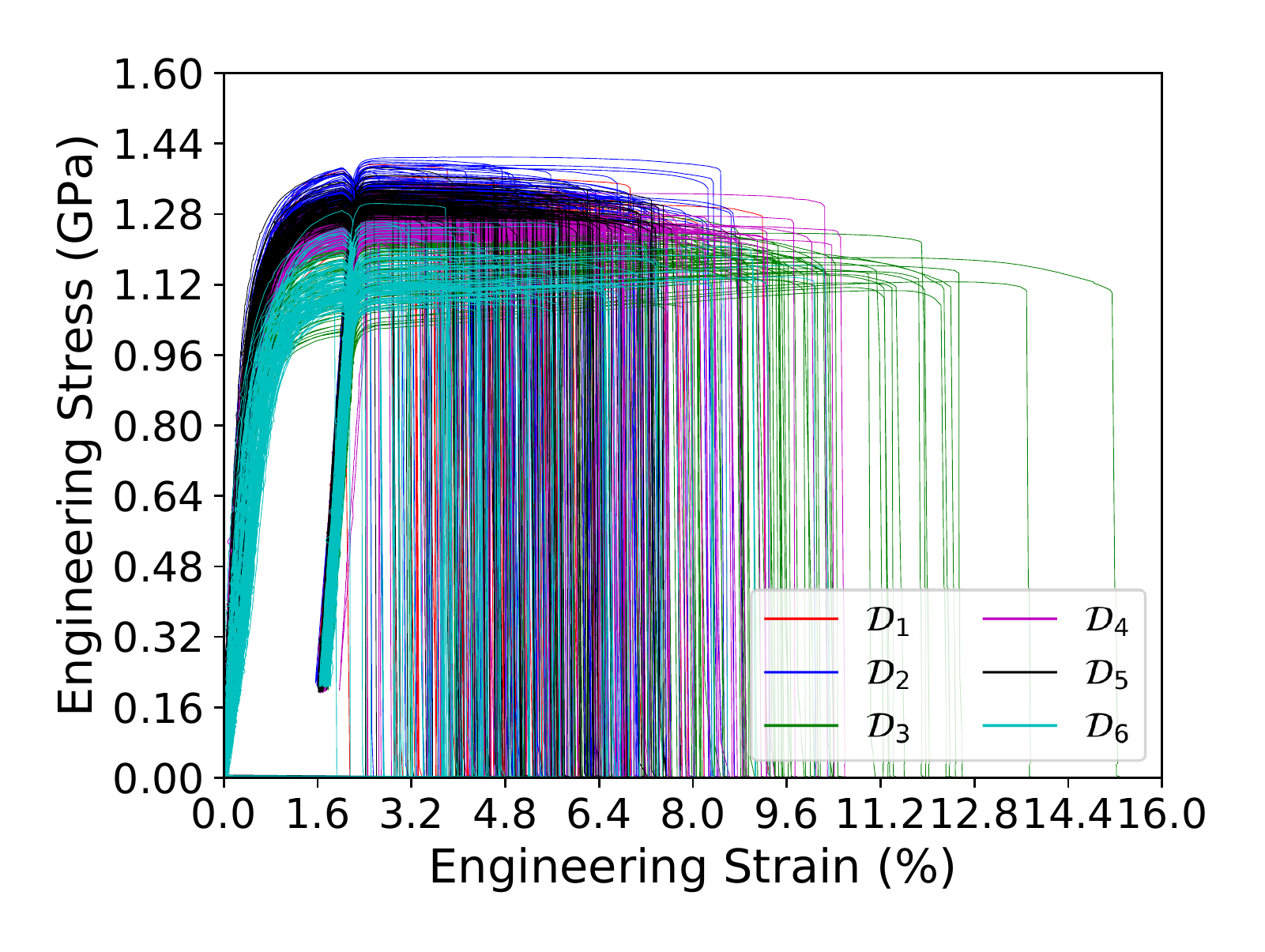}} \\
(b) \\
\end{tabular}
\begin{tabular}{c}
{\includegraphics[width=0.4\textwidth]{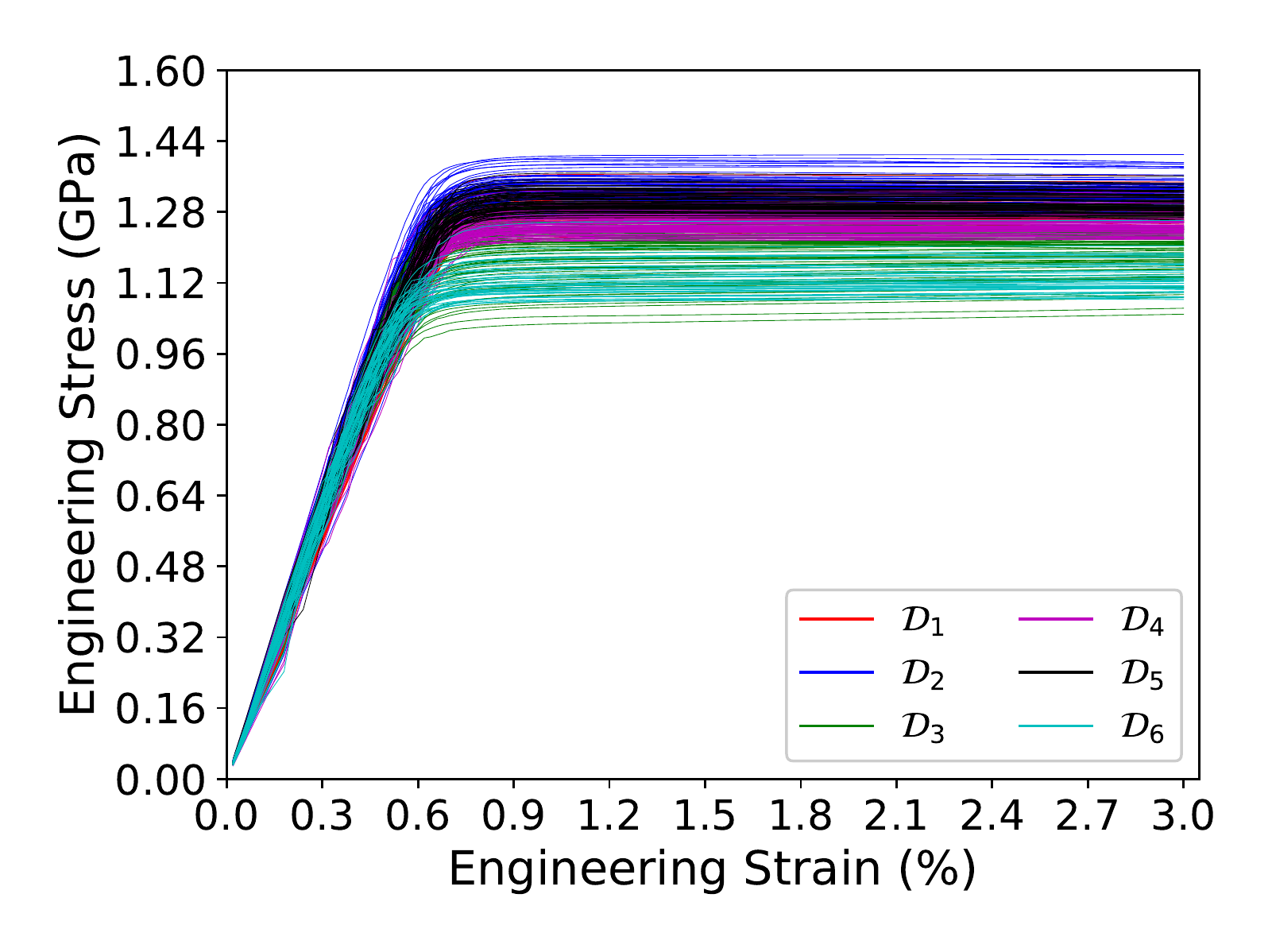}} \\
(c) \\
\end{tabular}}
\caption{(a) An array of nominally identical micro-tension ``dogbone'' specimens, 
(b) experimental data from Boyce \etal [\cite{boyce2017extreme}] color-coded by batch, 
and (c) the reduced data set used in this work.}
\label{fig:experiment}
\end{figure}

We focus this work on the analysis of high-throughput, micro-tension experimental measurements of additively manufactured stainless steel. 
From the experiments of Boyce \etal [\cite{boyce2017extreme}], we have six experimental data sets, each consisting of 120 stress-strain curves from the array of nominally identical dogbone-shaped specimens shown in \fref{fig:experiment}(a).
(The data from distinct builds of the array are referred to as \emph{batches} throughout the remainder of the manuscript.)
Each stress-strain curve, as shown in \fref{fig:experiment}(b), is qualitatively similar and behaves in a classically elastic-plastic fashion; however, the material displays a range of yield strengths, hardening and failure strengths and some variability in its apparent elastic properties.

To simplify the data and remove some of the uncertainties associated more with the loading apparatus than the material, we omit the pre-load cycle to approximately 0.2\% strain.
The remainder of the mechanical response is monotonic tensile loading at a constant strain rate, see \fref{fig:experiment}(c). 
We associate the zero strain reference configuration of each sample with the zero-stress, mildly worked material resulting from the pre-load cycle.
The resulting stress, $\stress$, and strain, $\strain$, values are derived from the customary engineering stress and strain formulas.
It is important to note that the both stress and strain are dependent on measurements of the specimen geometry.
The gauge section is nominally 1mm$\times$1mm$\times$4mm with variations of 2--10\%, depending on batch.
These variations are large relative to traditional manufacturing tolerances and reflect the current state-of-the-art in additive manufacturing process and quality control. 
Physically, the variations in dimensions and effective properties result from incomplete sintering and compaction of the particulate feed material and the subsequent porosity, surface roughness, residual stress and non-ideal microstructure.
We estimate the stress-strain measurement noise to be approximately Gaussian with $\pm$~0.009\% standard deviation in the strain measurement and $\pm$~20.0 MPa in the stress measurement based on the analysis of the random variations for individual curves and the noise in their zero-stress/zero-strain intercepts.
Since we do not try to model failure in this effort, we discard tests that do not reach at least 3\% strain.
This threshold was chosen to be sufficiently large that each sample curve is well within the plastic regime (and near peak stress), and yet retain sufficient data to enable calibration.
This pre-processing yielded $N_b = 6$ batches of stress data with $N_i = \{ 64, 77, 91, 79, 64, 46 \}$ curves, respectively.
To make the data suitable for the inverse problem of parameter calibration, we interpolate each curve and extract $n_{\strain}=151$ points over the interval (0,3)\% to finally arrive at
\begin{equation}\label{eq:data}
\Dc = \{ \Dc_{i} \}_{i=1}^{N_b}, \quad 
\text{with} \quad \Dc_i = \{ \Dc_i^{(k)}\}_{k=1}^{N_i} \quad 
\text{and} \ \Dc_i^{(k)} = \{ \stress^{(i,k)}_{j} \}_{j=0}^{n_{\strain}-1}, 
\end{equation}
where $i$ enumerates the batches, $k$ enumerates the $N_i$ stress curves within the $i$-th batch, and $\stress^{(i,k)}_{j} = \stress^{(i,k)}(\strain_j)$ represents the stress measured at the $j$-th strain value, $\strain_j = 0.03 j/(n_\strain-1)$, for the $k$-th curve of the $i$-th batch. 
The resulting data set is shown in \fref{fig:experiment}(c).

\section{Plasticity Theory} \label{sec:theory}

To model the observed behavior which resembles classical von Mises plastic response, we adopt a standard finite deformation framework [\cite{simo1988framework}] with a multiplicative decomposition of the deformation gradient, $\Fb$, into elastic and plastic parts
\begin{equation}\label{eq:W}
\Fb = \Fb_e \Fb_p \ ,
\end{equation}
where $\Fb_e$ is associated with lattice stretching and rotation, and $\Fb_p$ is associated with plastic flow.
Following \cite{Simo1998}, we assume an additive stored energy potential written in terms of the elastic deformation
\begin{equation}
W= \frac{\kappa}{2}\left(\frac{1}{2}(J_e^2-1)-\log(J_e)\right) + \frac{\mu}{2}\left(\tr[\bar{\bb}_e]-3\right).
\end{equation}
Here, the elastic volumetric deformation is given by $J_e=\det(\Fb_e)=\det(\Fb)$ since plastic flow is assumed to be isochoric, and the deviatoric elastic deformation is measured by $\bar{\bb}_e=J_e^{-2/3} \Fb_e\Fb_e^T$. 
We associate the elastic constants $\kappa$ and $\mu$ with the bulk modulus and shear modulus, respectively, and relate them to Young's modulus, $E$, and Poisson's ratio, $\nu$, via the linear elastic relations $\kappa=\frac{E}{3(1-2\nu)}$ and $\mu= \frac{E}{2(1+\nu)}$.
The Kirchhoff stress resulting from the derivative of the stored energy potential, $W$, is
\begin{equation} \label{eq:tau}
\taub = \frac{\kappa}{2}(J_e^2-1) \, \Ib + \sb \ \ \text{with} \ \
\sb=\mu \dev[\bar{\bb}_e].
\end{equation}

For the inelastic response, we employ a von Mises ($J_2$) yield condition between an effective stress derived from $\sb = \dev[\taub]$ and an associated flow stress, $\Upsilon$, as
\begin{equation} \label{eq:yield}
f:=\sqrt{\frac{2}{3}}\|\sb\|-\Upsilon \le 0.
\end{equation}
The rate independent, associative flow rule is written in the current configuration as the Lie derivative of the elastic left Cauchy-Green tensor (\cf \cite{Simo1998})
\begin{equation}
L_v\bb_e=-\frac{2}{3}\gamma~\tr[\bb_e]\frac{\sb}{\|\sb\|}.
\end{equation}
The Lagrange multiplier $\gamma$ enforces consistency of the plastic flow with the yield surface, obeys the usual Kuhn-Tucker conditions, and can be interpreted as the rate of plastic slip.
Finally, we make the flow stress 
\begin{equation} \label{eq:Y}
\Upsilon(\bar{\epsilon}_p) = Y + H\bar{\epsilon}_p + K(1-\exp(-B\bar{\epsilon}_p)),
\end{equation}
a function of the equivalent plastic strain
\begin{equation} \label{eq:eqps}
\bar{\epsilon}_p= \sqrt{\frac{2}{3}} \int_0^t \gamma dt,
\end{equation}
and the following parameters: initial yield, $\yield$; linear hardening coefficient, $\hard$, and nonlinear exponential saturation modulus $\satmod$ and exponent $\satexp$.
In tension, the yield strength, $\yield$, determines the onset of plasticity; the hardening coefficient $\hard$ determines the linear trend of the post-yield behavior; and $\satmod$, $\satexp$ superpose a more gradual transition in stress-strain from the trend determined by Young's modulus $\young$ in the elastic regime to $\hard$ in the plastic regime.
These material parameters form the basis of our analysis of material variability. 
To be clear, this standard J$_2$ plasticity model is a coarse-grained representation of the microstructural variations that engender the variability in the mechanical response, with the plastic strain covering a wide variety of underlying inelastic mechanisms and the physical definitions of the material parameters shaping our interpretation of the underlying causes of the variable response.

We approximate the tensile test with a boundary value problem on a rectangular parallelepiped of the nominal gauge section with prescribed displacements on two opposing faces and traction free conditions on the remaining faces to effect pure tension. 
Finite element simulations are performed in \textsc{Albany} [\cite{salinger2016albany}] using the constitutive model described in this section.
 The engineering stress $\stress$ and strain $\strain$ corresponding to that measured in the experiments are recovered from the reaction forces, prescribed displacements, cross-sectional area and gauge length.

\section{Calibration formulation} \label{sec:method}

In general, a calibration problem involves searching for the parameters $\thetab$ of a given model that minimize the difference between model predictions and observed data. 
In this work, we adopt a Bayesian approach to the calibration problem [\cite{kennedy2001bayesian,Sivia:1996,Rizzi:2012b,Rizzi:2013b,Sargsyan:2015,MARZOUK2007}].
In contrast to least-squares fitting resulting in a single set of parameter values, in a Bayesian perspective the parameters are considered random variables with associated probability density functions (PDFs) that incorporate both prior knowledge and measured data.  
The choice of Bayesian methods is well motivated by the data, which agree with the chosen model to a high degree but uncertainty is present in the model parameters both within and across all batches.
Bayesian calibration results in a joint posterior probability density of the parameters $\prob(\thetab|\Dc, M)$ that best fits the available observations $\Dc$ given the model choice $M$.
The parametric uncertainty reflected in the posterior PDF depends on the consistency of the model with the data and the amount of data. We aim to quantify the material variability using this probabilistic framework and physical interpretations of the parameters.

\subsection{Bayesian inference for parameter calibration} \label{sec:calibration}

Consider our model $M$ for the engineering stress $\stress=M(\strain;\thetab)$, being a full finite element plasticity model with an underlying plasticity model described in Section~\ref{sec:theory}, where $\strain$ is the independent variable and $\thetab$ is the vector containing physical parameters $\{ E, Y, H, K, B\}$ as well as auxiliary and nuisance parameters as will be defined later. By setting $\{H,K,B\}$ or $\{K,B\}$ to zero we can form a nested sequence of models with 2, 3, or 5 parameters with perfect plastic, linear hardening, or saturation hardening phenomenology, respectively. Given that we only have one dimensional tension data, we fix the Poisson's ratio $\nu=0.3$; however, we allow the Young's modulus, $E$, to vary, so that the locus of yield points is not constrained to a line.
We also allow for geometric variability through a non-dimensional cross-section correction factor, $A$, that influences the model output $\stress=A \times M(\strain;\thetab)$ and include $A$ in $\thetab$.
This geometric correction is motivated by the fact that the observed engineering stress data was computed using an average cross-sectional areas based on the outer dimensions of each sample, and can be interpreted as the ratio of the effective load bearing area of the sample to its measured average area.
The correction factor $A$ aims to mitigate the effect of utilizing one nominal value for cross-sectional area per sample rather than utilizing a more accurate spatially-varying area profile for each sample and the fact that the outer dimensions lead to an overestimate of actual load bearing area of the AM tensile specimens due to the physical imperfections discussed in \sref{sec:experiment}.
Also, since the gauge length, and hence the strain, are relatively error free, it is not included in the calibration parameters.

In a Bayesian setting for model calibration and selection, Bayes' rule is used to relate the information contained in the data and prior assumptions to the parameters in the form of a posterior probability density function as
\begin{equation} \label{eq:bayes}
\prob(\thetab|{\Dc}, M) 
= \frac{\prob({\Dc}|\thetab, M) \, \prob(\thetab|M)}{\prob(\Dc|M)} \ .
\end{equation}
Here $\prob({\Dc}|\thetab, M)$ is the likelihood of observing the data $\Dc$ given the parameters $\thetab$ and model $M$, $\prob(\thetab|M)$ is the prior density on the parameters reflecting our knowledge {\it before} incorporating the observations, and $\prob(\Dc|M)$ is the model evidence (which we will compute for model selection purposes). 
It is important to note that the denominator is typically ignored when sampling from the posterior since it is a normalizing factor, independent of $\thetab$, that ensures the posterior PDF to integrate to unity; however, this term, known as the model {\it evidence}, plays a central role in model selection, as will be described later.
In this context, we will employ relatively uninformative uniform prior densities due to lack of prior knowledge of the model parameters in the present context of the response of AM tensile specimens.
Experimental data influences the resulting posterior probability only through the likelihood $\prob({\Dc}|\thetab, M)$, which is based on some normalized measure of the distance between the data $\Dc$ and the model predictions $M(\strain; \thetab)$.
The likelihood plays an analogous role to the cost/objective function in traditional fitting/optimization in the sense that it describes the misfit between model predictions and observational data.
Specific forms of the likelihood will be discussed in \sref{sec:likelihood}.
As \eref{eq:bayes} suggests, the outcome is conditioned on the model chosen, leading to questions regarding model comparison and selection which will be discussed in \sref{sec:selection}.
In general, given the complexities of the model $M$, the posterior density $\prob(\thetab|{\Dc}, M)$ is not known in closed form and one has to resort to numerical methods to evaluate it. 
Markov chain Monte Carlo (MCMC) methods [\cite{gamerman2006markov,berg2008markov}] provide a suitable way to sample from the posterior density, while kernel density estimation, for example, can be used to provide subsequent estimates of the posterior PDF.

\subsection{Surrogate Modeling} \label{sec:surrogate}

The sampling of the parameters' posterior probability density and the evaluation of the Bayesian model evidence involves many evaluations of the computational model. 
Since the finite-element based forward model is relatively expensive to query (each tension simulation takes approximately 1 cpu-hour), the inverse problem of parameter estimation and model selection via direct evaluation becomes infeasible. 
Instead, we construct inexpensive-to-evaluate, accurate surrogates for the response of interest using polynomial chaos expansion (PCE)~[\cite{Wiener:1938,Ghanem:1991,Xiu:2002c}].
Marzouk {\it et al.}~[\cite{Marzouk:2007}] have shown that such surrogates can be effectively constructed using UQ techniques with a presumed uniform density on the parameters' range of interest.

Since rough bounds of each of the parameters can be estimated from the data and knowledge of similar materials, we represent the unknown parameters $\thetab$ using a spectral PCE in terms of a set of independent and identically distributed standard uniform random variables $\xib \sim [-1, 1]^{d_{\theta}}$, as in
\begin{equation}
\thetab(\xib) = \sum_{i=0}^{P_\theta} \thetab_{i} \Psi_i(\xib),
\label{eq:PCEpriorTheta}
\end{equation}
where $d_{\theta}$ represents the dimensionality of $\thetab$, $\Psi_I(\xib)$ are the orthogonal PC basis elements (Legendre polynomials in this case), and $P_\theta$ defines the number of terms in the expansion. 
\eref{eq:PCEpriorTheta} is, essentially, a linear transformation that maps standard uniform random variables to the unknown parameters over their range of interest.
A corresponding expansion of the model response, acting as a polynomial-based surrogate model, can be written as
\begin{equation} \label{eq:PCEforM}
M({\strain_j};\xib) \approx \sum_{i=0}^{P_M} \stress_i({\strain_j}) \Psi_i(\xib) \ .
\end{equation}
and is constructed as functions of the physical parameters, $\{ E, Y, H, K, B\}$ at each value of engineering strain $\strain_j$ at which data is available.
The PC coefficients for the inputs, $\thetab_{i}$, and outputs, $\stress_i({\strain_j})$, of the model  can be obtained, using one of two approaches: Galerkin projection or stochastic collocation. 
We utilize a non-intrusive stochastic collocation method with regression [\cite{xiu2010numerical}] to estimate the unknown PC coefficients as it does not require any modification to the existing computational models/simulators.
Details on this procedure are given in [\cite{OlmOmk:2010}], with an application of PCE surrogate modeling with computationally intensive numerical models in [\cite{Khalil:2015}]. 

\subsection{Likelihood Formulation} \label{sec:likelihood}

As mentioned, the likelihood is the term in Bayes' rule, \eref{eq:bayes}, that accounts for the data. 
Physical reasoning based on  data is available and its relationship with the model prediction helps to formulate the likelihood. 
From the discussion in \sref{sec:experiment}, one can argue that the batches are independent.
Furthermore, within a given batch, we assume all the $N_i$ stress-strain curves are independent since each experiment is a self-contained test, performed on separate specimens, \ie the variability of each specimen is the result of its specific microstructure.

We consider two different formulations of the likelihood, which lead to different formulations of the inverse problem, and hence models of the material variability.
The formulations differ by how they account for measurement noise and other variability, and how they are affected by (systematic) model discrepancies. 
Since each formulation leads to qualitatively different predictions, interpretations, and realizations, we are interested in how each is able to discriminate material variability from other sources of randomness.
In this section and in the Results section we will discuss how, given that plastic strain is a coarse metric of the inelastic deformation in additively manufactured materials, discrepancies between the observed data and the model predictions can be interpreted physically.
The results in \sref{sec:results} will illustrate how the posterior responds to the quantity of data and its variability.

\subsubsection{Additive error formulation} \label{sec:additive_error}

Consider the $k$-th stress-strain curve from the $i$-th batch which
consists of a sequence of stress observations $\{\stress_j^{(i,k)}\}_{j=0}^{n_\strain-1}$ obtained at the strain locations $\{\strain_j\}_{j=0}^{n_\strain-1}$. A widely-adopted approach is to express the discrepancy between an observation and surrogate model prediction using an additive noise model, as in
\begin{equation} \label{eq:model}
\stress_j^{(i,k)} = M(\strain_j;\thetab) + \eta_j^{(i,k)},
\end{equation}
where $\{\eta_j^{(i,k)}\}_{j=0}^{n_\strain-1}$ is the $(i,k)$ sample from the set of random variables $\{\eta_j\}_{j=0}^{n_\strain-1}$ capturing the discrepancy between observations and model predictions at a given $\strain_j$. 
This formulation is predicated on the assumption that the model $M(\strain;\thetab)$ \emph{accurately} represents the true, physical process occurring with fixed, but unknown, parameters. 
This a strong assumption (and one of the main deficiencies of this approach) since models are, in general, only approximations of observed behavior. 
Nevertheless, this is a commonly used method due to its simplicity.

In lieu of a completely characterized measurement error model (which is rarely obtained in practice), it is reasonable and expedient to assume the errors are distributed in an independent and identically distributed (i.i.d.) manner, with an assumed Gaussian probability density of zero mean (unbiased) and a variance parameterized by $\varsigma^2$. 
This yields the likelihood:
\begin{equation} \label{eq:additive_error}
\prob({\Dc_i^{(k)}}|\thetab, M) = 
\prod_{j=0}^{n_\strain-1} 
(2 \pi \varsigma^2)^{-1/2}
\exp\left( 
-\frac{(\stress_j^{(i,k)} - M(\strain_j;\thetab))^2}
{2 \varsigma^2}
\right),
\end{equation}
where we recall that $\Dc_i^{(k)}$ represents the stress observations collected from the $k$-th stress-strain curve of the $i$-th batch.
With the assumption that each experimental curve is independent from the others, the combined likelihood takes the form
\begin{equation} \label{eq:independent_additive_error}
\prob(\Dc_i|\thetab, M) = 
\prod_{j=0}^{n_\strain-1} \prod_{k=1}^{N_i}
(2 \pi \varsigma^2)^{-1/2}
\exp\left( 
-\frac{(\stress_j^{(i,k)} - M(\strain_j;\thetab))^2}
{2 \varsigma^2}
\right).
\end{equation}
for the full data set of the $i$-th batch.

The standard deviation $\varsigma$ can be either fixed in advance (based on prior knowledge) or inferred along with the other unknown parameters $\thetab$. 
Moreover, it can be assumed to be either constant or varying with the strain value. 
In this context, $\hat{\eta}$ represents the joint error attributed to modeling error as well as observational noise.
	
One can enrich this additive error formulation by adding a term capturing the discrepancy between the model prediction and truth (\ie  model error) separately.
A structure for the model error is more difficult to prescribe than that for the measurement error.
Given that this additional term is not physically associated with the presumed sources of non-measurement (material and other physical) variability, its applicability outside the training regime is tenuous.
Furthermore, the additive combination of the two error terms can lead to challenges in disambiguation in a parameter estimation context.
Lastly, this additive term can yield difficulties because it can lead to violations of physical laws and constraints [\cite{salloum2014inference,salloum2014inferenceb}].

\subsubsection{Embedded model discrepancy} \label{sec:embedded_discrepancy}

In order to avoid the difficulties associated with an additive error formulation, we embed the model error in key parameters, essentially converting the unknown parameters into random variables that introduce variability in model predictions due to their uncertain nature.
This approach, as detailed in [\cite{Sargsyan:2015}], represents selected parameters using PCE.
In our context, we will assume a uniform distribution for each parameter in \eref{eq:model}, given by the first-order Legendre-Uniform PCE:
\begin{equation} \label{eq:pcPerturbed}
\theta_i = \alpha_{i,0} + \alpha_{i,1} \xi_i \ ,
\end{equation}
in which $\alpha_{i,0}$ represents the mean term and $\alpha_{i,1}$ dictates the level of variability in $\theta_i$ which contributes to the total model error.
We will also consider the model selection problem as to whether or not to embed the error in $\theta_i$, which amounts to keeping the $\alpha_{i,1}$ term or setting it explicitly to zero.
These coefficients (one or two per parameter) need to be inferred from the experimental observations.
Such embedding of error, along with a classical additive error, may account for both model and measurement errors, respectively.
A noticeable advantage of this approach, when compared to employing additive error alone, is that the model error is embedded in the model parameters and hence we can propagate the calibrated model error through numerical simulations to any output of interest.
(Note that this use of a PCE is distinct from its use in the surrogate modeling, where the extent $\thetab$ was selected to cover the feasible range of the parameters and not inferred, as it is in this context.)

The problem of calibrating such an embedded error model is equivalent to that of estimating the probability density of the parameters in which the error is embedded. 
Specifically, our objective is to estimate $\alphab=\{\alpha_{i,0}, \alpha_{i,1}, \ldots\}$ that parametrize the density of $\thetab$.
This is in contrast to the conventional use of Bayesian inference for parameter estimation, \ie additive error formulations, in which one infers the \emph{parameter} and not its density.
Also, the data for our present calibration problem motivates the embedded approach since it suggests the uncertainties are aleatory/irreducible rather than epistemic/reducible.

In this context, the model calibration problem thus involves finding the posterior distribution on $\alphab$ via Bayes' theorem \eref{eq:bayes}
\begin{equation} \label{eq:bayes_embedded}
\prob(\alphab|{\Dc}, M) = \frac{\prob({\Dc}|\alphab, M) \, \prob(\alphab|M)}{\prob({\Dc} | M)},
\end{equation}
where $\alphab$ has been substituted for $\thetab$,
$\prob(\alphab|{\Dc}, M)$ denotes the posterior PDF, $\prob(\Dc|\alphab, M)$ is the likelihood PDF, and $\prob(\alphab|M)$ is the prior PDF.
Note that $\alphab$ reduces to the classical parameter vector $\thetab$ when no error embedding is performed ($\theta_i = \alpha_{i,0}$ and $\alpha_{i,j} = 0$ for $j>0$).
Among the different options detailed in [\cite{Sargsyan:2015}] for the likelihood construction, we employ the marginalized likelihood, which for the $i$-th batch $\Dc_i$, can be written as
\begin{align}
\prob({\Dc_i}|\alphab, M) = 
\frac{1}{(2\pi)^{\frac{N_i n_\strain}{2}}}
\prod_{j=0}^{n_\strain-1}
\prod_{k=1}^{N_i} 
\frac{1}{\varsigma_{j}(\alphab)}
\exp\left[-\frac{(\mu_j(\alphab)-\stress_j^{(i,k)})^2}
{2\varsigma_{j}^2(\alphab)}\right],
\end{align}
where
\begin{align}
\mu_j(\alphab) = 
\mathbb{E}_{\xib}[M(\strain_j;\thetab(\alphab,\xib))]
\end{align}
and
\begin{align}
\varsigma_j^2(\alphab) = 
\mathbb{V}_{\xib}[M(\strain_j;\thetab(\alphab,\xib))] + \varsigma^2
\end{align}
are the mean and variance of the model predicted stress at fixed $\alphab$ and strain point $\strain_j$.
These moments are computed using the quadrature techniques that are commonly relied upon in uncertainty propagation, as in the input characterization, \eref{eq:pcPerturbed}; see [\cite{Sargsyan:2015}] for detailed description of the methodology involved in likelihood evaluation.

\subsection{Bayesian model selection} \label{sec:selection}

The expansion in \eref{eq:pcPerturbed} resulting from the embedded model discrepancy approach is general in that it does not impose limitations on which parameters should embed the error.
Furthermore, as mentioned in \sref{sec:calibration}, we can model the stress using 2, 3, or 5 physical parameters.
We have also chosen to examine the inclusion of a cross-section correction factor, $A$, which could also be chosen as another parameter in which to embed an error.
Therefore, there are three physical models that are competing to fit the data, with up to 6 physical parameters, all of which are competing to be bestowed with an embedded error term.
That amounts to a total of 88 plausible models that are competing to fit the given data.
Since determining the optimal model and optimal parameters for model error embedding a priori is not possible, we perform model selection using Bayes factors [\cite{Berger:1996a,Verdinelli:1996}].
This method of model selection is a data-based approach that selects the optimal model as the one that strikes the right balance between data-fit and model simplicity [\cite{Beck2010a,Kass1995}].
This balance is monitored using the so-called model evidence, or marginal likelihood, for each model $M$, given by
\begin{equation} \label{eq:mod_evid}
\prob({\Dc} | M) = \int \prob({\Dc}|\alphab, M) \, \prob(\alphab|M) \, {\rm d} \alphab \ .
\end{equation}

The computation of the model evidence $\prob({\Dc} | M)$ is typically neglected in Bayesian calibration as it acts merely as a normalizing factor in Bayes' rule, \eref{eq:bayes}. 
The model evidence acts as a quantitative Ockham's razor that, when maximized, performs an explicit trade-off between the data-fit and the model simplicity [\cite{Beck2010a}]. This is elucidated when examining the logarithm of the model evidence [\cite{Muto:2008, Sandhu:2017}]:
\begin{equation} \label{eq:ockham}
\ln \prob(\Dc | M) = \mathbb{E}[\ln \prob({\Dc}|\alphab, M)] - \mathbb{E}\left[\ln \frac{\prob(\alphab|{\Dc},M)}{\prob(\alphab|M)}\right]
\end{equation}
where the expectation $\mathbb{E}[\cdot]$ is with respect to the posterior PDF $\prob(\alphab|{\Dc},M)$.
The first term on the right hand side of \eref{eq:ockham} quantifies the data-fit and is known as {\it goodness-of-fit}.
The second term is equal to the relative entropy (or Kullback-Leibler divergence) between the prior and posterior PDFs, also known as the {\it information gain} [\cite{Konishi:2007}].
Given sufficient data $\Dc$, the information gain is normally higher for more complex models (models with more parameters or with parameters of greater prior PDF support reflecting poor prior knowledge as to their values).
Hence, the information gain term quantifies model complexity by examining the relative difference between prior and posterior PDFs.
This is in contrast to the frequentist approach where model complexity, in general, only depends on the number of model parameters rather than their relative prior supports [\cite{Konishi:2007}].

Model selection using Bayes factors involves the direct comparison of two plausible models at a time, with Bayes factor defined as the ratio of their evidences given by
\begin{eqnarray}
B(M_i,M_j)= \frac{\prob(\Dc | M_i)}{\prob(\Dc | M_j)} = \frac{\int \prob({\Dc}|\alphab, M_i) \, \prob(\alphab|M_i) \, {\rm d} \alphab}{\int \prob({\Dc}|\alphab, M_j) \, \prob(\alphab|M_j) \, {\rm d} \alphab}
\label{eq:Bayes_factor}
\end{eqnarray}
with $B(M_i,M_j) > 1$ indicating that model $M_i$ is more likely while $B(M_i,M_j) < 1$ indicates that $M_j$ is the more likely model.
One can supplement the evidence with prior belief on these models (called prior model probability) and in that case we would be interested in the ratios of posterior model probabilities.
In our context, we assume that all models have equal prior probabilities ahead of analyzing the data and thus Bayes factor is a suitable way to compare the plausible models.

The key challenge is to compute the model evidence efficiently and accurately. In this investigation, we will modify a technique known as adaptive Gauss-Hermite quadrature [\cite{Naylor:1982,Liu:1994,Hakim:2018,Khalil:2018}] by employing importance sampling for integration rather than Gauss quadrature. In an importance sampling framework, we
utilize a multivariate normal distribution $q$ as a proposal/sampling distribution, with a mean vector $\boldsymbol{\mu}$ and covariance matrix $\Sigma$ equal to the posterior mean and posterior covariance of $\alphab$, i.e.
\begin{eqnarray}
q(\alphab) =\frac{1}{\sqrt{(2\pi)^{d_\alpha}|\Sigma|}}\exp\left(-\frac{1}{2}\left(\alphab-\boldsymbol{\mu}\right)^\intercal\Sigma^{-1}\left(\alphab-\boldsymbol{\mu}\right)\right).
\label{eq:weight_function}
\end{eqnarray}

The model evidence $\evidence$, being an integral over the parameter space, can now be estimated using importance sampling with $q$ as the proposal distribution. To do this, we reformulate the evidence as
\begin{eqnarray}
\evidence = \int \prob({\Dc}|\alphab, M) \, \prob(\alphab|M) \, {\rm d} \alphab
= \int \frac{\int \prob({\Dc}|\alphab, M) \, \prob(\alphab|M) }{q(\alphab)}
q(\alphab) {\rm d} \alphab \ ,
\label{eq:evidence}
\end{eqnarray}
and utilize the change of variable $\alphab =L\widetilde{\alphab}+\boldsymbol{\mu}$, with $L$ being the lower triangular matrix in the Cholesky decomposition of the covariance matrix, $\Sigma = LL^\intercal$, to obtain the following formulation for the evidence:
\begin{eqnarray}
\evidence = \int \frac{\int \prob({\Dc}|L\widetilde{\alphab}+\boldsymbol{\mu}, M) \, \prob(L\widetilde{\alphab}+\boldsymbol{\mu}|M) }{\exp\left(-\frac{1}{2}\widetilde{\alphab}^\intercal\widetilde{\alphab}\right)}
\exp\left(-\frac{1}{2}\widetilde{\alphab}^\intercal\widetilde{\alphab}\right) \, |L| \, {\rm d} \widetilde{\alphab} \ .
\label{eq:adaptive_quad}
\end{eqnarray}

$\evidence$ can now be approximated using Monte Carlo sampling strategy as
\begin{eqnarray}
\evidence \approx \frac{1}{N_{LH}} \sum_{k=1}^{N_{LH}} \vert L\vert \frac{\int \prob({\Dc}|L\widetilde{\alphab_k}+\boldsymbol{\mu}, M) \, \prob(L\widetilde{\alphab_k}+\boldsymbol{\mu}|M) }{\exp\left(-\frac{1}{2}\widetilde{\alphab_k}^\intercal\widetilde{\alphab_k}\right)},
\label{eq:imp_samp}
\end{eqnarray}
where $(\widetilde{\boldsymbol{\alpha_k}})_{1\le k \le N_{LH}}$ are Latin Hypercube samples [\cite{OlmOmk:2010}] drawn from a standard multi-variate normal distribution Gauss-Hermite quadrature points
where $N_{LH}$. For the estimator in \eref{eq:imp_samp} to be accurate, we need good estimates for the posterior mean vector and covariance vector, $\boldsymbol{\mu}$ and $\Sigma$), which can be obtained adaptively using the following iterations ($n$ denoting the iteration number):
\begin{eqnarray}
\boldsymbol{\mu}_n &=& \frac{1}{N_{LH}} \sum_{k=1}^{N_{LH}} \left[ \begin{split} \vert L_{n-1}\vert \frac{\prob({\Dc}|L\widetilde{\alphab_k}+\boldsymbol{\mu_{n-1}},M) p(L_{n-1}\widetilde{\boldsymbol{\alpha_k}}+\boldsymbol{\mu}_{n-1}|M)}{\exp\left(-\frac{1}{2}\widetilde{\boldsymbol{\alpha_k}}^\intercal \widetilde{\boldsymbol{\alpha_k}}\right)} \ \times \ \\
\\
& \hspace*{-80mm} (L_{n-1}\widetilde{\boldsymbol{\alpha_k}}+\boldsymbol{\mu}_{n-1}) \end{split}\right]\\
\Sigma_n &=&  \frac{1}{N_{LH}} \sum_{k=1}^{N_{LH}} \left[ \begin{split} \vert L_{n-1}\vert \frac{\prob({\Dc}|L\widetilde{\alphab_k}+\boldsymbol{\mu_{n-1}},M)p(L_{n-1}\widetilde{\boldsymbol{\alpha_k}}+\boldsymbol{\mu}_{n-1}|M)}{\exp\left(-\frac{1}{2}\widetilde{\boldsymbol{\alpha_k}}^\intercal \widetilde{\boldsymbol{\alpha_k}}\right)} \ \times \ \\
\\
& \hspace*{-80mm} (L_{n-1}\widetilde{\boldsymbol{\alpha_k}})(L_{n-1}\widetilde{\boldsymbol{\alpha_k}})^\intercal\end{split}\right]\\
\evidence_n &=& \frac{1}{N_{LH}} \sum_{k=1}^{N_{LH}} \vert L_n\vert \frac{\prob({\Dc}|L\widetilde{\alphab_k}+\boldsymbol{\mu_{n}},M)p(L_n\widetilde{\boldsymbol{\alpha_k}}+\boldsymbol{\mu}_n|M)}{\exp\left(-\frac{1}{2}\widetilde{\boldsymbol{\alpha_k}}^\intercal \widetilde{\boldsymbol{\alpha_k}}\right)}
\label{eq:iteration}
\end{eqnarray}

For each individual model, these iterations start with an initial guess for the posterior parameter mean vector and covariance matrix. We obtain those from an initial run of a Markov chain Monte Carlo sampler over the parameter posterior. The iterations stop with a termination criterion on $\evidence$. For this application, the termination criteria is met when the relative change in $\evidence$ is less than $10^{-8}$.

\section{Results} \label{sec:results}

In this section, we present results relating to the construction of surrogate models from the full finite element plasticity model, their calibration to the experimental data, and the selection of the optimal predictive model using Bayesian model selection.
Most of the results are obtained using the UQ Toolkit [\cite{uqtk:web}].

\subsection{Surrogate Model Construction} \label{sec:surrogate_construction}

To describe the material stress-strain behavior we analyze, calibrate and compare three nested plasticity models of increasingly complex phenomenology, namely perfect plasticity, linear hardening and saturation hardening. As mentioned in \sref{sec:theory}, we focus on five parameters: Young's modulus, $\young$; yield strength, $\yield$; hardening modulus, $\hard$; saturation modulus, $\satmod$; and saturation exponent, $\satexp$, which control the elastic-plastic stress response.
We construct surrogate model for this five parameter model, as outlined in \sref{sec:surrogate}. In order to construct Legendre-Uniform PCE surrogates for the model outputs (engineering stresses at a collection of engineering strain values), we assumed that the material parameters are uniformly distributed and characterized by the following first-order PCE:
\begin{alignat}{5} 
\young  &=& 200  \  &+& 80     \,\xi_1  \quad &\text{[GPa]}, \nonumber \\
\yield  &=& 1.2  \  &+& 0.5    \,\xi_2  \quad &\text{[GPa]}, \nonumber \\
\hard   &=& 3.0  \  &+& 3.0    \,\xi_3  \quad &\text{[GPa]}, \label{eq:inputPCE} \\
\satmod &=& 0.2  \  &+& 0.2    \,\xi_4  \quad &\text{[GPa]}, \nonumber \\
\satexp &=& 1750 \  &+& 1750   \,\xi_5  \quad &  \nonumber
\end{alignat}
where $\{\xi_1,\xi_2,\xi_3,\xi_4,\xi_5\} \sim {\cal U}(-1,1)$ are independent identically distributed (i.i.d.) standard uniform random variables.
The geometric parameter $A$ will be discussed in the next section.

We chose these parameters ranges to be large enough that the corresponding predictions can capture the variability of the experimental data shown in \fref{fig:experiment}b.
The ranges are also chosen so that the constructed surrogate can be used to emulate the two and three parameter model by setting the relevant parameters to zero.
Also, we remark that the expansion with i.i.d.~random variables is a common step to build the surrogate model.
Any correlations between the parameters will then be discovered through the inverse problem, see [\cite{MARZOUK2007,Rizzi:2012b,Rizzi:2013b,Khalil:2015,Safta:2017}] for example.
To construct and validate this surrogate, we collected $5032$ training and $500$ validation samples.

\begin{figure}[!t]
\centering
{\includegraphics[width=0.55\textwidth]{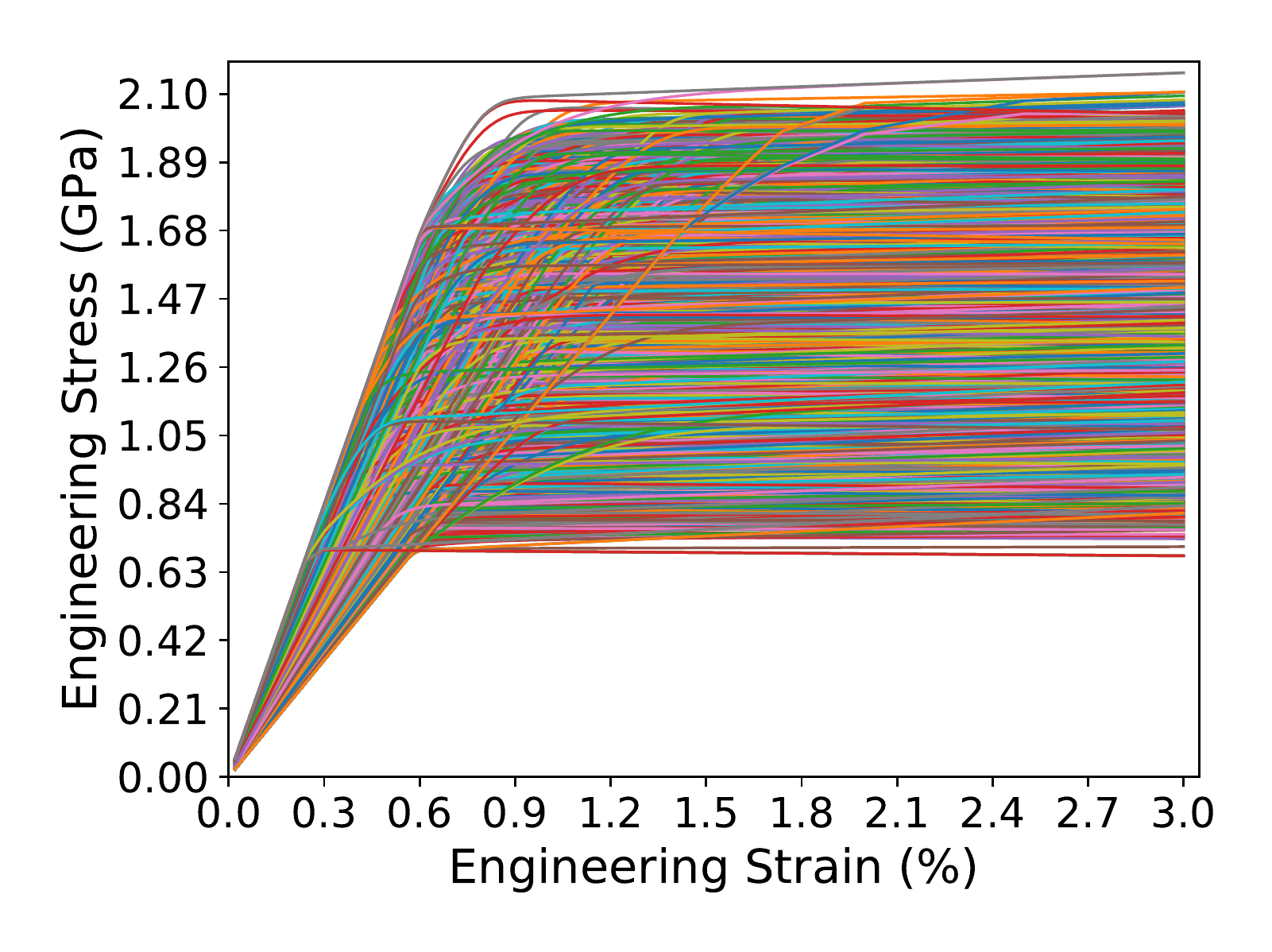}}
\caption{Stress-strain curves samples used to build the surrogate for the five parameter model.
} \label{fig:5P_samples}
\end{figure}

\begin{figure}[!t]
\centering
{\includegraphics[width=\figwidth]{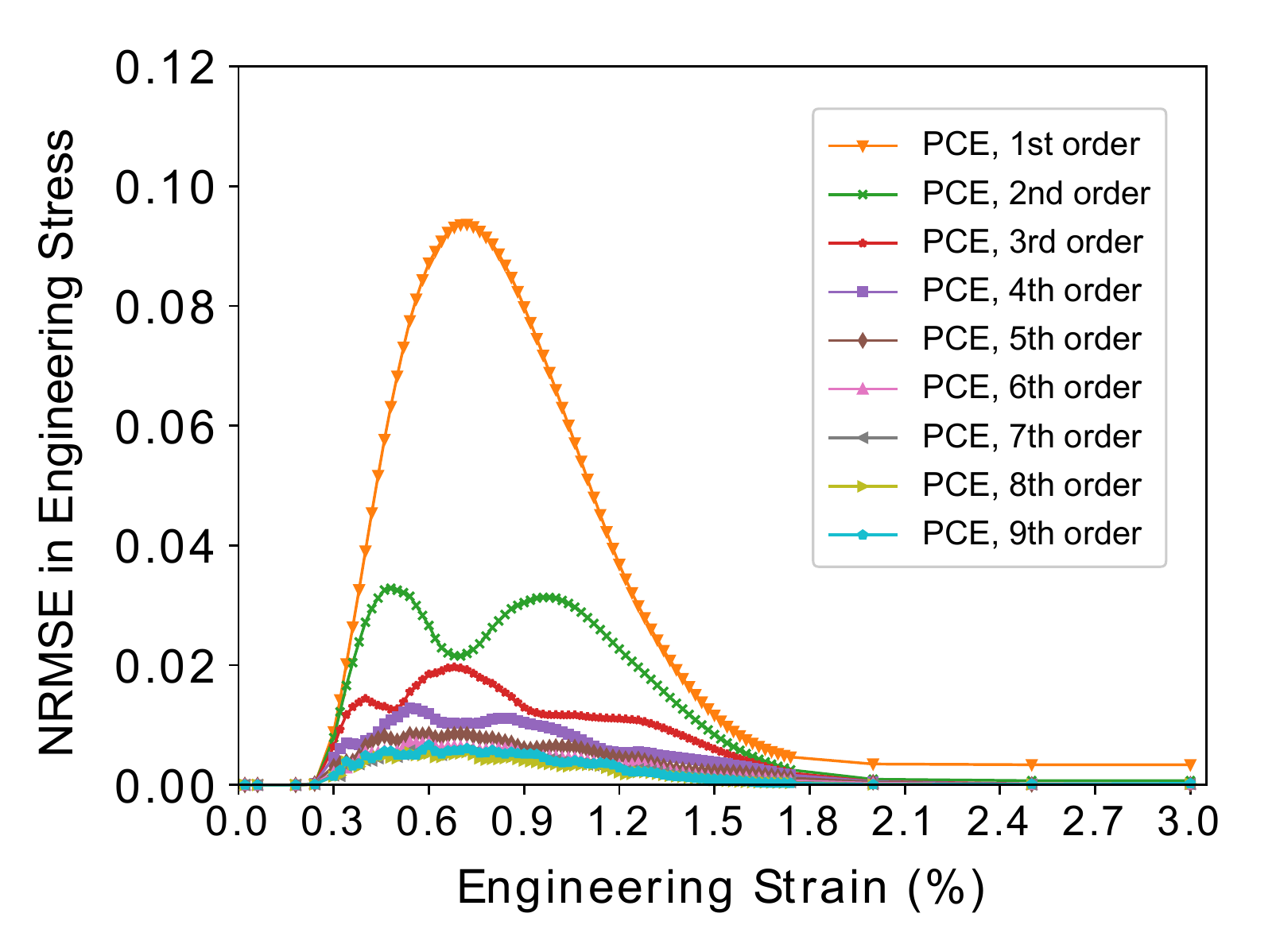}}
\caption{Normalized root-mean-square error in the 5-parameter PCE surrogate for the engineering stress as a function of engineering strain. Results are shown for PCE surrogates of orders up to nine.
} \label{fig:5P_errors}
\end{figure}

\fref{fig:5P_samples} shows the stress-strain curves obtained from the corresponding plasticity simulations over the 5032 points in parameter space.
\fref{fig:5P_errors} shows the normalized (relative) root-mean-square error for the various PCE surrogates with increasing order as a function of the strain. In this case, a PCE surrogate of order 6 or greater exhibits the lowest errors. Since the errors do not drop significantly for PCE surrogates of order beyond 6, we are likely to be over-fitting the full simulation data when using regression in fitting the coefficients of higher order PCE models.
For the 2 (perfect plasticity) and 3 (linear hardening) parameter models, a piece-wise surrogate was superior to a global representation due to the lower smoothness of response relative to the 5-parameter model.
The piece-wise surrogate was constructed using plastic strain as a classifier to switch between surrogates for the elastic and plastic regimes.

\subsection{Model selection using Bayes' factor} \label{sec:mod_sel_res}

For illustrative purposes, we will use the third batch of available data as described in \sref{sec:experiment}. We will reserve the joint analysis of all available batches for possible future investigations due to the observed large and abrupt changes in stress-strain behaviors across batches (as opposed to a more smooth variability within batches) as illustrated in \fref{fig:experiment}c. As was discussed in \sref{sec:selection}, the stress can be modeled using 2, 3, or 5 physical parameters, resulting in three physical models that are competing to fit the data. Furthermore, depending on the number of model being analyzed, there are up to 6 parameters (including the cross-section correction factor) to be bestowed with an embedded error term. That amounts to a total of 88 plausible models that are competing to fit the given data.

We apply the methodology outlined in \sref{sec:selection} to compete the model evidence of each of these models.
The competing models are labeled according to (a) the inclusion/exclusion of physical parameters and (b) whether or not an error is embedded in those physical parameters.
For clarity, models will be distinguished by subscript labels.
The appearance of the parameter in the label subscript indicates its inclusion in the model, while the appearance of the parameter with a $\delta$ subscript denotes the inclusion of the parameter, as well as an embedded error as in \eref{eq:pcPerturbed}.
For example, model $M_{E_\delta Y_\delta H K B A}$ is a 5-parameter plasticity model with errors embedded in Young's modulus and yield strength whereas $M_{E Y A_\delta}$ is a 2-parameter plasticity model with an error embedded in the cross-section correction factor.
For all models, we assume the measurement noise intensity (or standard deviation), $\varsigma$, to be an unknown constant along the strain axis, \ie the measurement error does not depend on the observed strain nor stress values.
We therefore infer $\varsigma$ along with the other parameters.
To enforce positivity of $\varsigma$ in the inference step, we chose to infer the natural logarithm of $\varsigma$ with a uniform prior on [-5,0] in log($\varsigma$).
For the material property parameters, we choose uniform priors with ranges coinciding with those chosen to build the surrogate model in \eref{sec:surrogate_construction}.
As for the cross-sectional area correction factor, $A$, we utilized a uniform prior with support on [0,2].

Having computed the model evidence for all 88 plausible models, it was observed that the model evidence was maximized for model $M_{E Y_\delta H K B A_\delta}$ and thus we compute Bayes factor with respect to this model as reported in \tref{tab:bayes_factor}.
We can see that model $M_{E Y_\delta H K B A_\delta}$ is the most likely model (based on model evidence), with the second-most likely model, $M_{E Y_\delta H K B_\delta A_\delta}$, being 40\% less likely.

The following observations are noteworthy:
\begin{itemize}
	\item The 5 parameter model is the most likely physical model, in comparison to the 3 and 2 parameter models (as described in \sref{sec:theory}).
	\item The variability in the mechanical responses as observed from experiments can be effectively and compactly modeled via an embedded error in the yield strength $Y$ and the cross-sectional area correction factor $A$. 
Physically, this result is plausible given the dimensional variations and the fact that the AM process is likely to affect plastic response and unlikely to alter the effective elastic modulus.
	\item The data provides relatively weak evidence to suggest that the variability can be modeled by embedding the errors in the hardening modulus, $H$, saturation exponent $B$, or Young's modulus, $E$. 
This too is plausible given the small variations in hardening and the trend in near-yield behavior, refer to \fref{fig:experiment}c.
\end{itemize}

\begin{table}
	\begin{center}
		\begin{tabular}{c|c}
			\hline
			\hline
			Model $M_i$ & Bayes Factor with respect to $M_{E Y_\delta H K B A_\delta}$ \\ \hline
			$M_{E Y_\delta H K B A_\delta}$			&		1\\
			$M_{E Y_\delta H K B_\delta A_\delta}$					&		0.60\\
			$M_{E Y_\delta H_\delta K B A_\delta}$					&		0.14\\
			$M_{E_\delta Y_\delta H K B_\delta A_\delta}$					&		0.05\\
			$M_{E Y_\delta H K_\delta B_\delta A_\delta}$					&		0.02\\
			$M_{E_\delta Y_\delta H K B A_\delta}$					&		0.01\\
			Other models  &  $< 0.01$ \\
			\hline
		\end{tabular}
	\end{center}
	\caption{Bayes factor $B(M_i,...)$ as defined in Eq.~(\ref{eq:Bayes_factor}) with respect to $M_{E Y_\delta H K B A_\delta}$}
	\label{tab:bayes_factor}
\end{table}

\subsection{Calibration} \label{sec:calibration_results}

\begin{figure}[h!]
\centering
{\includegraphics[width=0.9\textwidth]{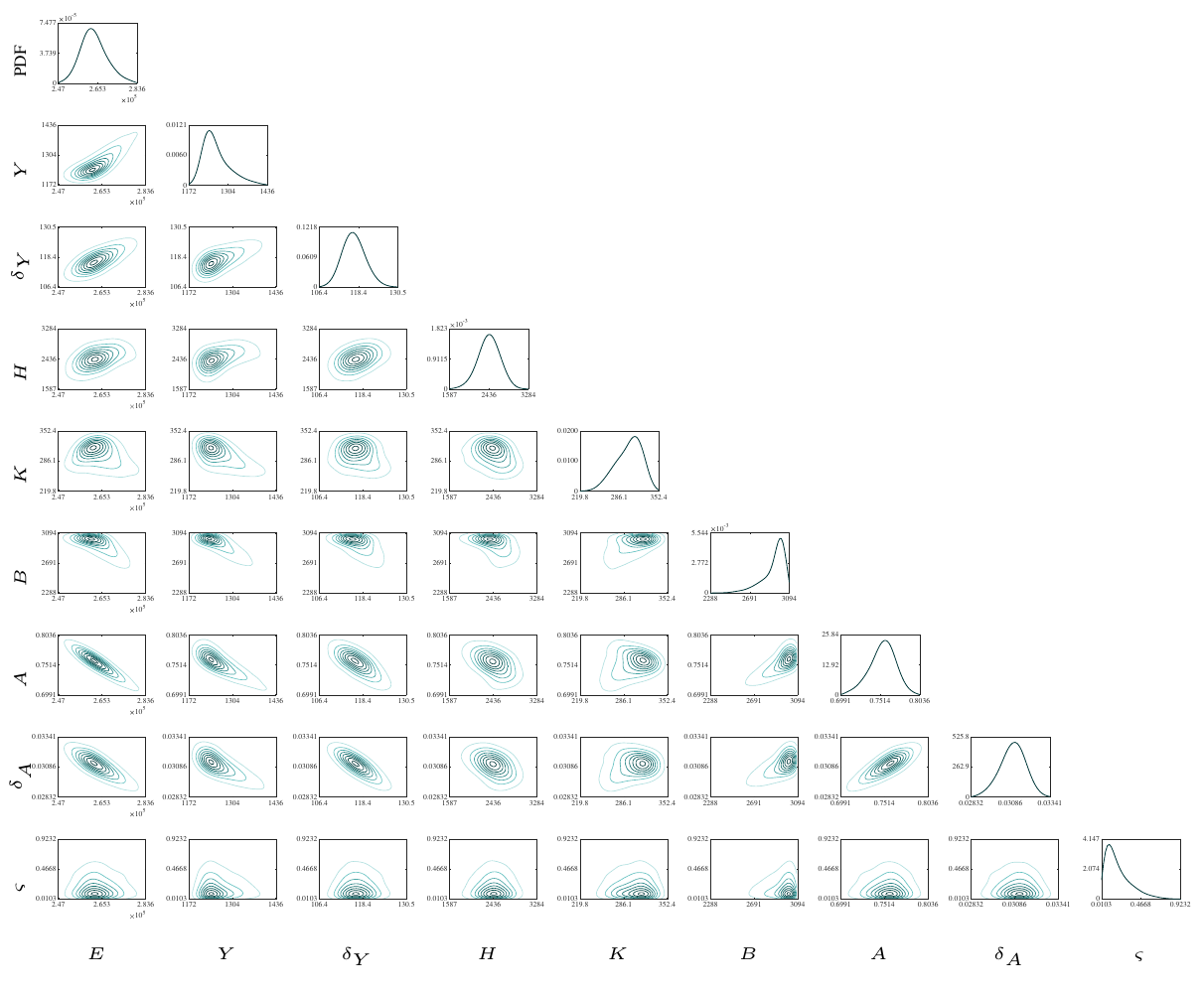}}
\caption{1D and 2D marginals of the posterior parameter PDF for model $M_{E Y_\delta H K B A_\delta}$.}
\label{fig:posterior_pdfs}
\end{figure}

The previous analysis has resulted in an optimal model $M_{E Y_\delta H K B A_\delta}$ that strikes a balance between data-misfit and model simplicity (as indexed by the number of parameters and which are allowed to be distributions).
As part of the previous analysis, we sampled from the posterior parameter PDF using adaptive Metropolis-Hastings Markov chain Monte Carlo sampling technique [\cite{gamerman2006markov,berg2008markov}].
Figure~\ref{fig:posterior_pdfs} shows the 1D and 2D marginal posterior parameter PDFs, in which $\delta_Y$ and $\delta_A$ denote the PCE coefficients that account for the embedded errors in the respective parameters.
Note the unimodality of the PDFs, implying the existence of a global maximum-a-posteriori (MAP) estimate for the parameters.
There are additional insights that can be made based on further examination of the (marginal) PDFs in \fref{fig:posterior_pdfs}:
\begin{itemize}
	\item The cross-sectional area correction factor $A$ has a mean value of 0.76 and varies with most probability between 0.7 and 0.8.
This indicates that the post-processing that was carried out in converting the applied loads to engineering stresses relied upon over-estimates of the sample cross-sectional areas, as expected given that the effective load bearing area is at most the area given by the outer dimensions of the gauge section.
	\item There is strong negative correlation between Young's modulus, $E$, and the cross-sectional area correction factor, $A$.
This is to be expected given the basic relationship between force, area and stress and, also, the fact that the stress at any strain value is bounded within the envelope of variability.
	\item There is strong negative correlation between the intensity of the embedded errors in yield strength, $\delta_Y$, and the cross-sectional area correction factor, $\delta_A$.
This is logical since the data suggests a fixed level of variability in the engineering stress at any strain value and both of these terms combine in contributing to that total variability.
	\item Examining the marginal PDF of the intensity of the additive error, $\varsigma$, we observe a relatively insignificant MAP estimate close to 0.02 GPa. This indicates that the embedded error terms are effective in capturing the variability in the mechanical response, with an embedded error in the yield strength having a MAP estimate close to 115 GPa (from the marginal posterior PDF of $\delta_Y$).
\end{itemize}

\begin{figure}[h!]
\centering
{\setlength{\tabcolsep}{0mm} \begin{tabular}{cc}
{\includegraphics[width=0.45\textwidth]{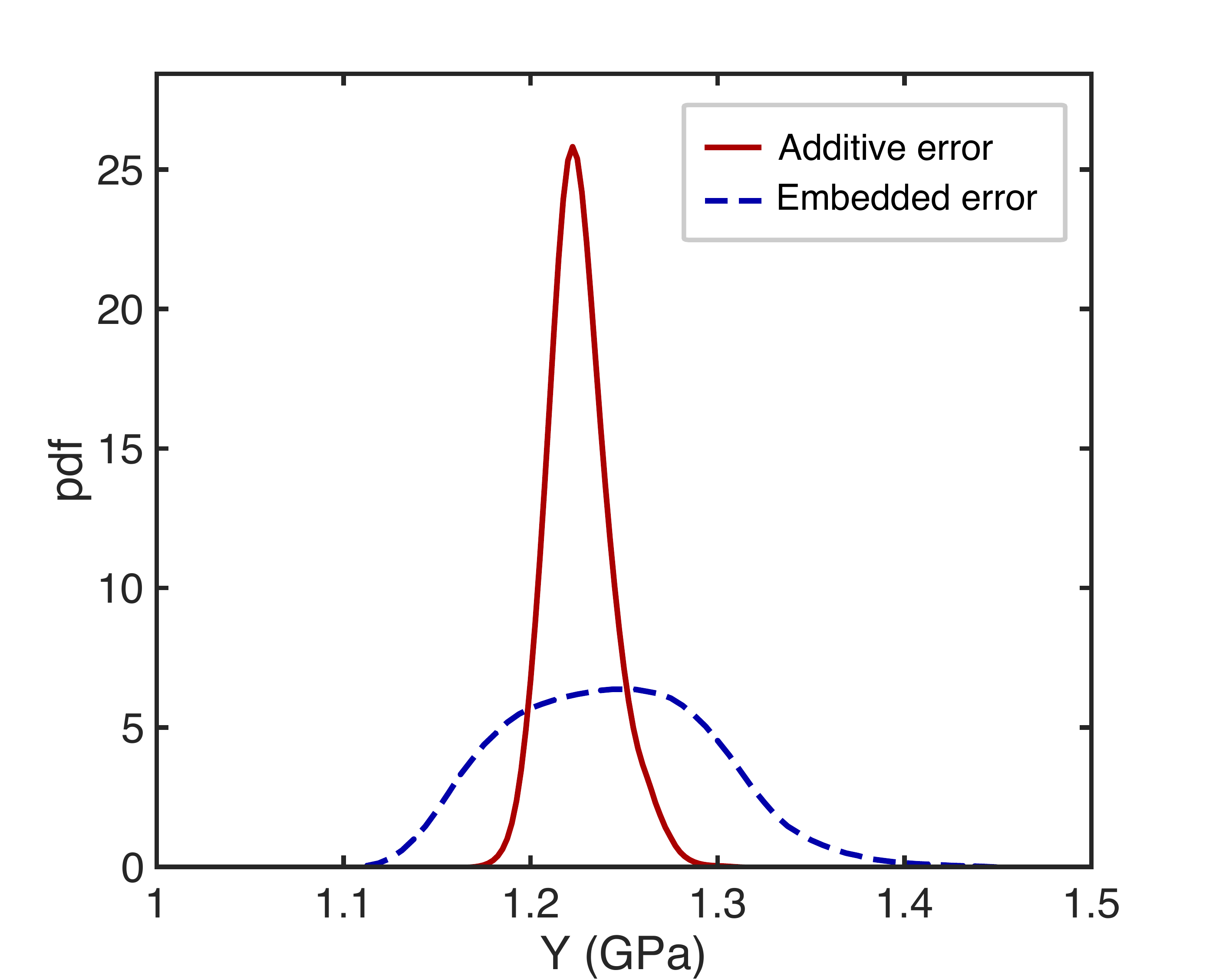}} & 
{\includegraphics[width=0.45\textwidth]{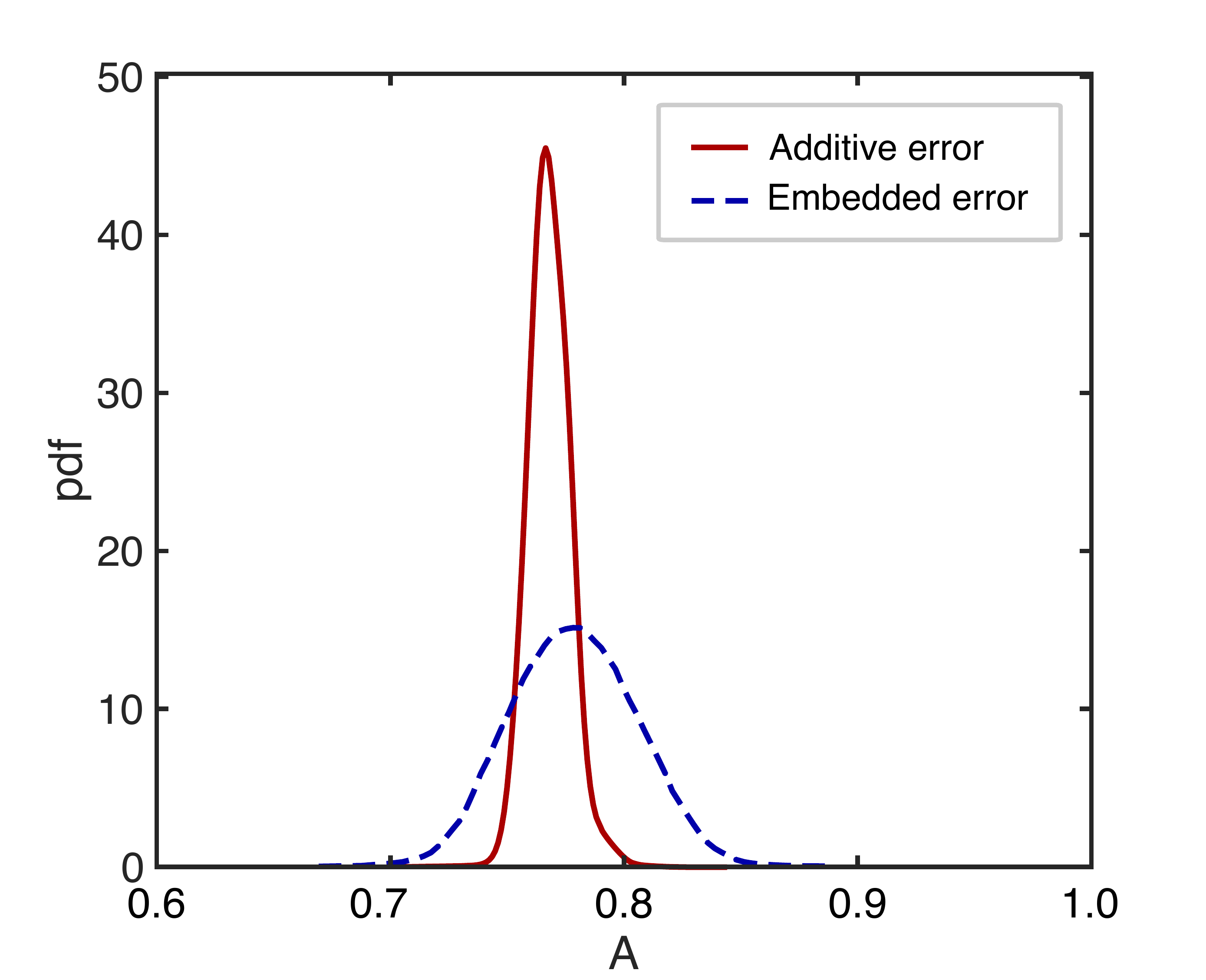}} \\
			(a) & (b) \\
\end{tabular}}
\caption{Comparison of posterior PDFs obtained using the additive (red, solid curves) and embedded (blue, dashed curves) error models for the parameters (a) $Y$ yield strength and (b) area correction factor $A$.}
\label{fig:cmpclassEmbe}
\end{figure}

We next examine the differences in the inferred posterior parameter PDFs between the additive and embedded error models.
To reiterate, the key difference between the two approaches are the embedded errors which explicitly allow the calibrated models to reflect the inherent variability in the mechanical responses using parametric variability.
In contrast, calibration with solely an additive error attributes variability in the observed quantities (i.e. engineering stress) to an additive error term which is not connected directly to the variability of any particular parameter.
We focus on the two parameters that are selected to carry an embedded error in the optimal model according to the model selection study (see \sref{sec:mod_sel_res}).
Those parameters are the yield strength, $\yield$, and cross-sectional area correction factor, $A$.
\fref{fig:cmpclassEmbe} shows the posterior PDFs for the two parameters obtained using the additive and embedded error approach.
We highlight the following key observations:
\begin{itemize}
	\item It is obvious that the embedded approach attributes a significant amount of variability in yield strength $Y$ based on the data while the additive approach results in smaller uncertainty.
	In effect, the distribution in the additive approach converges to a delta distribution at the best average value for $Y$, since the parameter is treated as an unknown constant rather than a random variable.
The width of the distribution in the additive approach characterizes the lack of data (not the inherent material variability), while the embedded approach converges to a (non-degenerate) distribution characterizing the range of $Y$ that help explain the observed variability.
	\item Likewise, the difference in the two approaches is readily apparent in the distributions of the cross-sectional area correction factor $A$ shown in \figref{fig:cmpclassEmbe}.
	Although the mean estimate for $A$ is more or less the same using the two approaches, the uncertainty (reflected in the support of the PDFs) is larger for the embedded error approach and is interpretable as the variation of a physical parameter responsible for the inconsistent mechanical response.
\end{itemize}

\subsection{Predictions under uncertainty} \label{sec:pred_results}

\begin{figure}[h!]
\centering
{\setlength{\tabcolsep}{0mm} \begin{tabular}{cc}
{\includegraphics[width=0.45\textwidth]{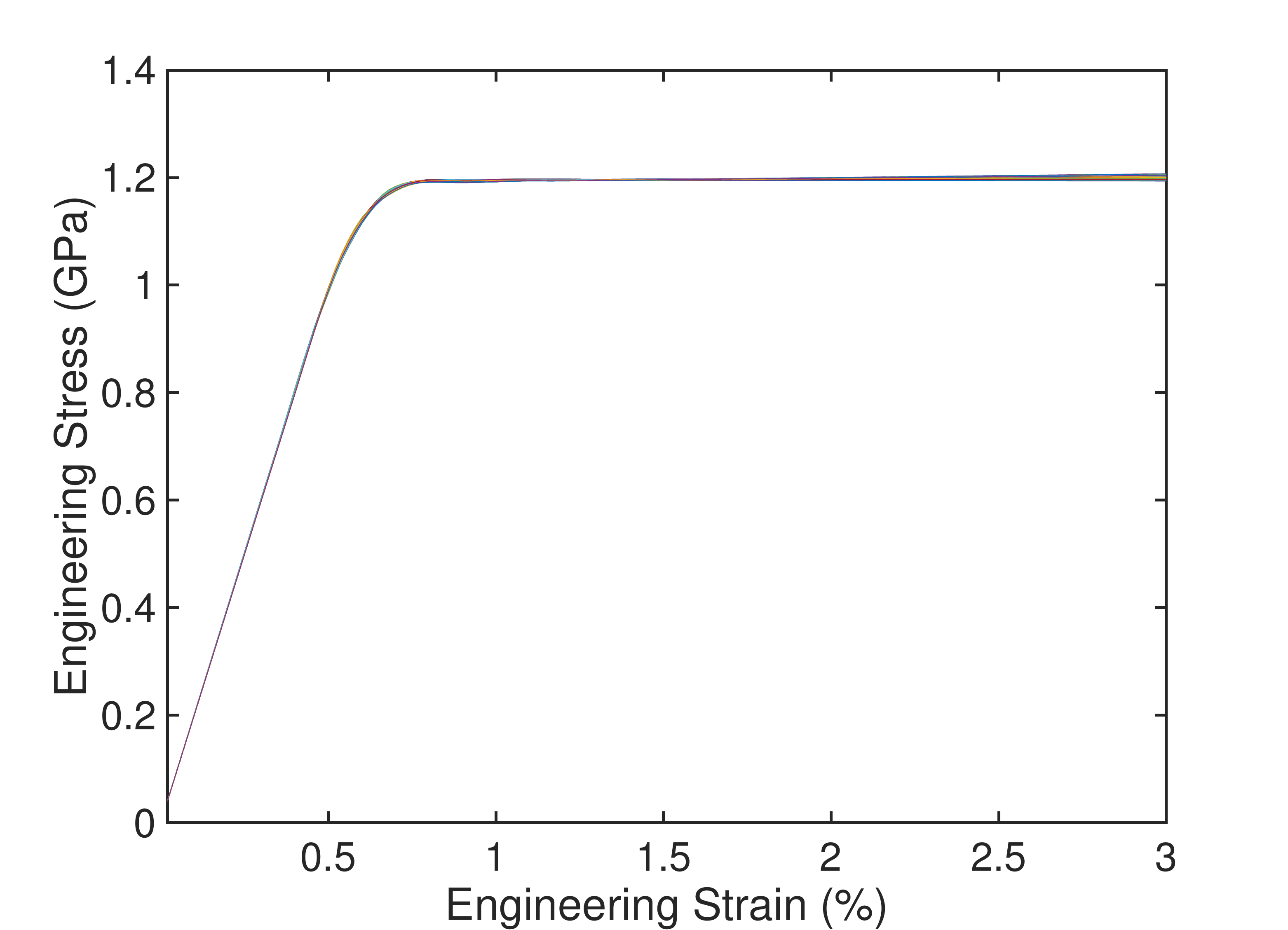}} & 
{\includegraphics[width=0.45\textwidth]{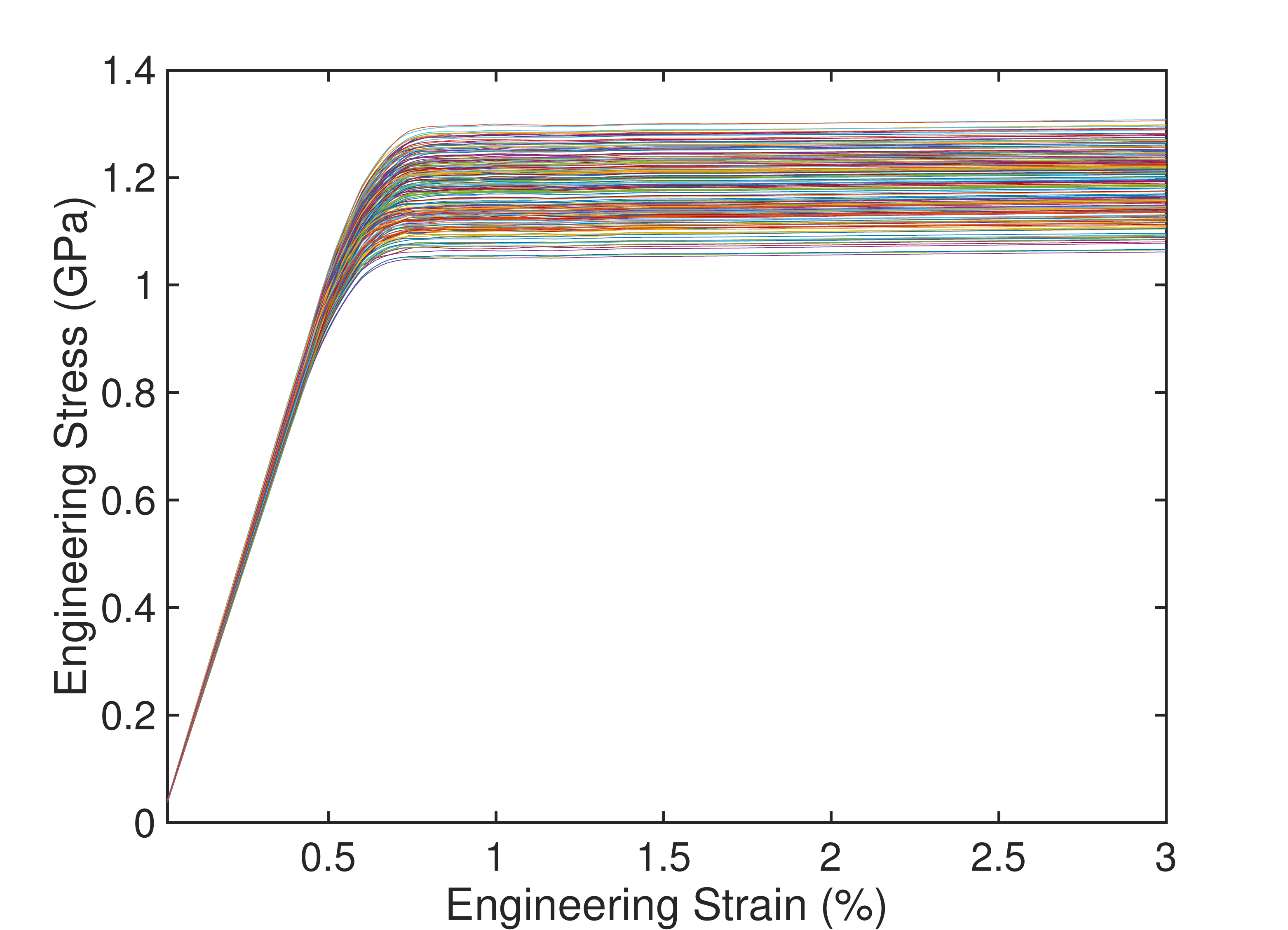}} \\
			(a) & (b) \\
{\includegraphics[width=0.45\textwidth]{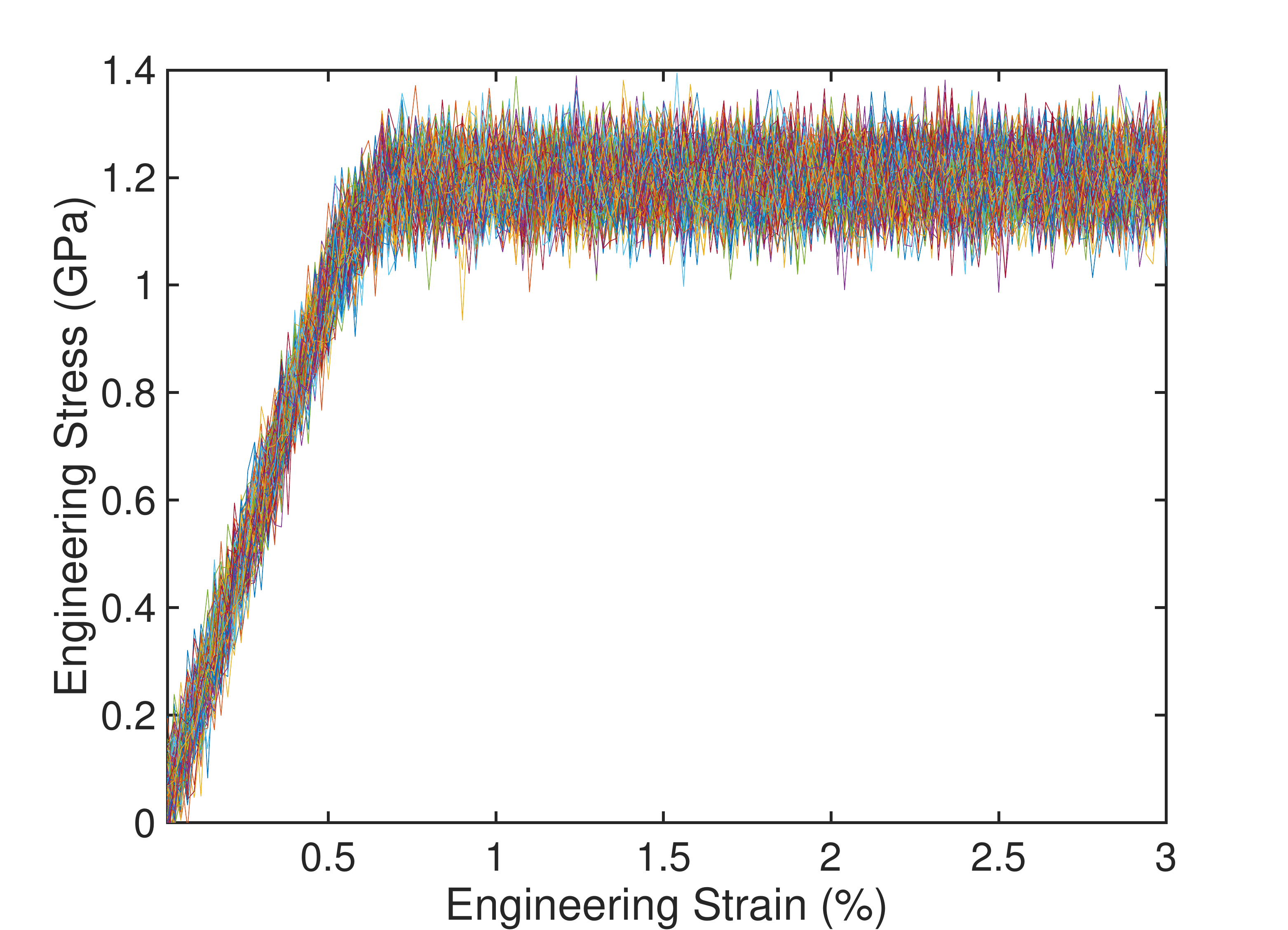}} & 
{\includegraphics[width=0.45\textwidth]{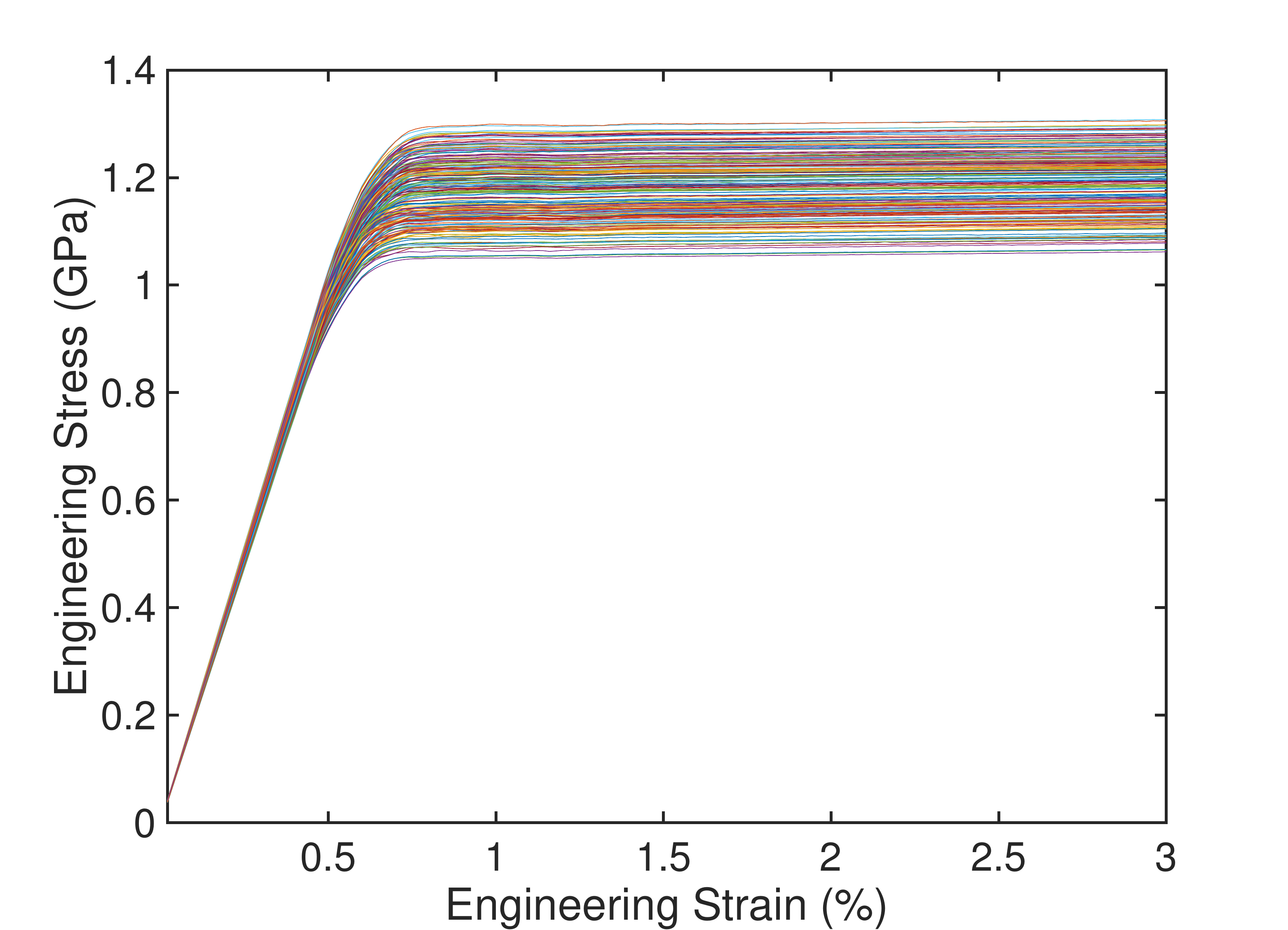}} \\
			(c) & (d) \\
\end{tabular}}
{\setlength{\tabcolsep}{0mm} \begin{tabular}{c}
{\includegraphics[width=0.45\textwidth]{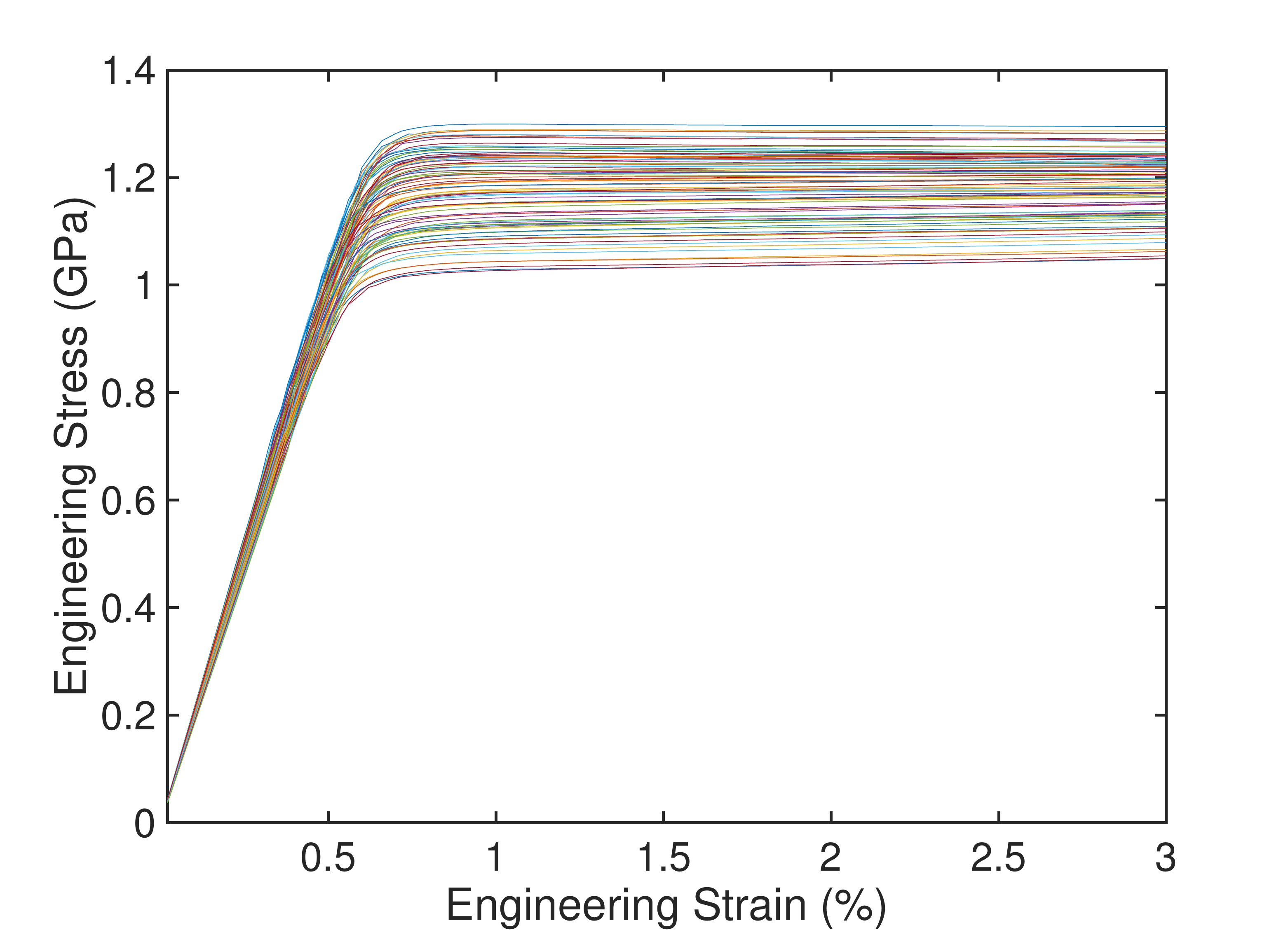}} \\
(e) \\
\end{tabular}}
\caption{Sample posterior predictive realizations obtained using the 5 parameter model for the additive (a,c) and  embedded (b,d) approach. The top row shows the results obtained by sampling the joint parameter posterior density and propagating these samples through the model only, while the middle row shows the results with the contribution of the additive error. The experimental observations are shown in (e) for comparison.}
\label{fig:realizPlots}
\end{figure}

In this section, we will examine the predictions obtained using the optimal model (see previous model selection sections) with embedded model error in yield strength, $\delta_Y$, and the cross-sectional area correction factor, $\delta_A$.
We also examine the predictions obtained using the 5-parameter model with a purely classical additive model error, $M_{E Y H K B A}$. 
We obtain 150 posterior predictive realizations of the stress-strain curves from both models as shown in \fref{fig:realizPlots} reflecting the uncertainty in the calibrated parameters and the embedded error terms (for model $M_{E Y_\delta H K B A_\delta}$).
The top row shows the results obtained by sampling the joint posterior parameter PDF of the parameters $\alphab$, and pushing these samples through the model only; while the middle row shows the results with the contribution of the additive error term. 
These results highlight the differences between the additive and embedded error approach in characterizing material variability effects on mechanical responses:
\begin{itemize}
        \item It is evident from comparing \fref{fig:realizPlots}a,c (additive error) and \fref{fig:realizPlots}b,d (embedded error) to the data shown in \fref{fig:realizPlots}e that the embedded error approach yields a suitable representation where the variability of the response is determined by the variability of the material parameters (with the aid of embedded parametric distributions).
	\item The additive approach yields a tight envelope of predictions, and the full variability is only captured by the additive error term, which results in over-estimation of the predictive uncertainty.
	\item Comparing individual realizations to the curves obtained experimentally, it is clear that the embedded approach (with and without additive error contribution) yields curves that match quite well the trends observed in the experiment by appropriately characterizing material variability and measurement noise, while the classical method deviates considerably and is dominated by the high frequency, uncorrelated white noise.
\end{itemize}

\section{Conclusions} \label{sec:conclusion}

We have presented a method which can model the variability of a material which is well-described by existing plasticity models of mean response, but contains microstructural variability leading to different macroscopic material properties.
By leveraging the embedded error method of Sargsyan \etal [\cite{sargsyan2015statistical}] (which, as mentioned, was originally developed to address model discrepancy), we can  mathematically represent the variability in mechanical response as material and geometric parameters drawn from a well-calibrated joint distribution.
This Bayesian approach was shown to be consistent with our understanding of microstructural variability and allows physical interpretation of the sources of variability through the parameters.
These interpretations can be linked to processing through inference and intuition.
The Bayesian methodology is appropriate for UQ studies requiring forward propagation of physical variability and also enables the utilization of Bayesian model selection to arrive at an optimal model that strikes a balance between data-misfit and model complexity.
The constitutive model we developed is amenable to non-intrusive sampling (and adaptable to direct evaluation in simulation codes that handle fields of distributions), and thus enables a robust design methodology that can predict performance margins with high confidence.

Another important contribution of this work is the illustration of the contrast the proposed approach with commonly used uncertainty formulations.
The standard, additive error formulation appropriately accounts for the uncertainty in the experiments arising from measurement error.
Yet, in the case of inherent variability, it characterizes all the uncertainty as measurement error which results in unwarranted confidence in the material properties and an inability to correctly and physically understand how that variability would manifest in applications.
We have demonstrated that the embedded method accurately characterizes the aleatory uncertainty present in the experimental observations and enables engineering UQ analysis of material and geometric sources of uncertainty.
It gives insight into what aspects of a homogeneous, macro-scale constitutive model are most strongly affected by microstructural variability and enables quantitative model selection.
Moreover, it is capable of representing both significant external noise and inherent variability in a unified formulation.

In future work, we will extend the methodology to the post-necking failure behavior of similar materials.

\section*{Acknowledgments}
This work was supported by the LDRD program at Sandia National Laboratories, and its support is gratefully acknowledged.
B.L. Boyce would like to acknowledge the support of the Center for Integrated Nanotechnologies.
Sandia National Laboratories is a multimission laboratory managed and operated by National Technology and Engineering Solutions of Sandia, LLC., a wholly owned subsidiary of Honeywell International, Inc., for the U.S. Department of Energy's National Nuclear Security Administration under contract DE-NA0003525.


\end{document}